\documentclass[a4paper,11pt]{article}

\pdfoutput=1 
\usepackage{jcappub} 
\usepackage{eurosym}

\usepackage[T1]{fontenc} 
\usepackage{amssymb}
\usepackage[dvipsnames]{xcolor}
\usepackage[normalem]{ulem}
\usepackage{systeme}
\usepackage{bm}
\usepackage{verbatim}
\usepackage{amsmath}
\usepackage[makeroom]{cancel}
\usepackage{hyperref}

\newcommand{\be}{\begin{equation}}
\newcommand{\ee}{\end{equation}}
\newcommand{\bea}{\begin{eqnarray}}
\newcommand{\eea}{\end{eqnarray}}

\newcommand{\ylm}[1]{Y_{\ell m}( #1)}

\def\d{{\rm d}}
\def\H{{\mathcal H}}
\def\n{\boldsymbol{n}}
\def\k{\boldsymbol{k}}
\def\ve{\boldsymbol{v}}
\def\alm{a_{\ell m}}

\def\odm{{\Omega_{\rm cdm  0}}}
\def\ob{{\Omega_{\rm b 0}}}

\def\fnl{{f_{\rm NL}}}
\def\HI{\rm HI}

\title{Multi-wavelength spectroscopic probes: prospects for primordial non-Gaussianity and relativistic effects}

\author[1]{Jan-Albert Viljoen,}
\author[2,1]{Jos\'e Fonseca,}
\author[1,3]{Roy Maartens}
\affiliation[1]{Department of Physics \& Astronomy, University of the Western Cape,\\Cape Town 7535, South Africa}
\affiliation[2]{School of Physics \& Astronomy, Queen Mary University of London, London E1 4NS, UK}
\affiliation[3]{Institute of Cosmology \& Gravitation, University of Portsmouth,  Portsmouth PO1 3FX, UK}

\emailAdd{javiljoen74@gmail.com}

\abstract{Next-generation cosmological surveys will observe larger cosmic volumes than ever before, enabling us to access information on the primordial Universe, as well as on relativistic effects. We consider forthcoming 21cm intensity mapping surveys (SKAO) and optical galaxy surveys (DESI and Euclid), combining the information via multi-tracer cross-correlations that suppress cosmic variance on ultra-large scales. 
In order to fully incorporate wide-angle effects and redshift-bin cross-correlations, together with lensing magnification and other relativistic effects, we use the angular power spectra, $C_\ell(z_i,z_j)$. Applying a Fisher analysis, we forecast the expected precision on $\fnl$ and the detectability of lensing and other relativistic effects. We find that the full combination of two pairs of 21cm and galaxy surveys, one pair at low redshift and one at high redshift, could deliver $\sigma(\fnl)\sim 1.5$, detect the Doppler effect with a signal-to-noise ratio $\sim$8 and measure the lensing convergence contribution at $\sim$2\% precision. In a companion paper, we show that the best-fit values of $\fnl$ and of standard cosmological parameters are significantly biased if the lensing contribution neglected.}

\begin{document}
\maketitle
\flushbottom

\section{Introduction}

Measuring the effects of primordial non-Gaussianity in the large-scale structure is an important scientific driver of future surveys as their volume increases. Primordial non-Gaussianity of the different types (local, equilateral, folded) provides an exquisite window into the physics of the very early universe (see e.g. \cite{Bartolo:2004if} for a review). These types of primordial non-Gaussianity have been constrained by the Planck survey \citep{Akrami:2019izv}, which is the current state-of-the-art.

Non-gaussianities give rise to  non-zero  odd-point clustering functions, but local-type primordial non-Gaussianity  also affects the two-point statistics of tracers of the dark matter  \citep{Matarrese:2008nc,Dalal:2007cu}. It induces a scale-dependent correction to the bias, $\propto \fnl H^2/k^2$ in Fourier space, thereby affecting the power spectrum on very large scales. The state-of-the-art constraint on the local-type primordial non-Gaussianity parameter $\fnl$  is provided by Planck \citep{Akrami:2019izv}: $f_{\rm NL} = -0.9 \pm 5.1$ (at 68\% CL). Several constraints have also been provided by quasar surveys  \cite{Leistedt:2014zqa,Castorina:2019wmr},  the most recent  being $f_{\rm NL} = -12 \pm 21$  \cite{Mueller:2021tqa}.

Future galaxy surveys will improve on current constraints because they will access many more long modes than current surveys \cite{Giannantonio:2011ya,Camera:2013kpa,Font-Ribera:2013rwa,Camera:2014bwa,Alonso:2015uua,Raccanelli:2015vla,MoradinezhadDizgah:2018zrs,Fonseca:2020lmi}. Measuring the scale-dependent bias {using the power spectrum} on very large scales requires extremely large cosmological volumes to reduce error bars; in most cases this is not enough due to the growth of cosmic variance with scale. In fact, {forecasts that are based only on the power spectrum for realistic future cosmological surveys, using the scale-dependent bias of a single tracer on its own, cannot} provide a precision of $\sigma(\fnl)<1$. This is an important threshold to distinguish between single-field and many multi-field inflationary scenarios (see e.g. \cite{dePutter:2016trg}).
In order to beat cosmic variance and approach or break through the $\sigma(\fnl)\sim 1$ target, one needs to take advantage of the multi-tracer technique \cite{Seljak:2008xr,McDonald:2008sh,Hamaus:2011dq,Abramo:2013awa,Abramo:2015iga},  as shown in \cite{Ferramacho:2014pua, Yamauchi:2014ioa,Alonso:2015sfa,Fonseca:2015laa,Fonseca:2016xvi,Fonseca:2018hsu,Bacon:2018dui,Gomes:2019ejy,Ballardini:2019wxj,Bermejo-Climent:2021jxf}.

It is important to note that the observed power spectrum includes light-cone effects \cite{Yoo:2012se,Challinor:2011bk,Bonvin:2011bg,Alonso:2015uua}, which can produce a signal similar to that of $\fnl$, potentially leading to a bias in the measurement of $\fnl$ \cite{Bruni:2011ta,Jeong:2011as,Namikawa:2011yr,Camera:2014sba,Alonso:2015uua,Lorenz:2017iez,Bernal:2020pwq}. (This is also the case for the bispectrum \cite{Kehagias:2015tda,DiDio:2016gpd,Koyama:2018ttg,Maartens:2020jzf}.) Other theoretical and systematic effects may also be degenerate with $\fnl$ in spectroscopic surveys (see e.g. \cite{Castorina:2021xzs,Taylor:2021bhg,Matthewson:2021rmb}).  We include large-scale lensing magnification and all other light-cone  effects, and we show in an accompanying paper \cite{Viljoen:2021ocx} that neglecting these relativistic effects in a multi-tracer analysis can bias estimates of $\fnl$ and of the standard cosmological parameters (see also \cite{Bernal:2020pwq}). However, lensing and other relativistic effects are not simply a theoretical `nuisance' that needs to be included in the modelling -- they can also be important cosmological probes of gravity (see e.g. \cite{Bonvin:2013ogt,Montanari:2015rga,Hall:2016bmm,Bonvin:2016dze,Lepori:2017twd,Andrianomena:2018aad,Bonvin:2018ckp,Clarkson:2018dwn,Franco:2019wbj,deWeerd:2019cae,Maartens:2019yhx,Beutler:2020evf,Jolicoeur:2020eup,Umeh:2020cag}).

In this paper, we study combinations of future large-scale structure surveys in optical/ NIR and radio bands, in order to forecast how well we will be able to detect relativistic effects and constrain $\fnl$.  One of our goals is to also determine how well we could detect the Doppler term, both in single and multiple tracer cases, which requires accurate redshifts. We therefore use `spectroscopic' to mean cosmological surveys with high redshift accuracy, irrespective of how such redshifts are effectively measured.  While photometric surveys provide lower shot noise, which is an important advantage in the multi-tracer technique, they also lose velocity information via averaging over thick redshift bins. We, therefore, consider surveys with specifications similar to the following planned surveys:  the Dark Energy Spectroscopic Instrument (DESI) Bright Galaxy Sample (BGS) \cite{2016arXiv161100036D};  the Euclid spectroscopic H$\alpha$ survey \cite{Euclid:2019clj};  the 21cm intensity mapping surveys in Bands 1 and 2 with the MID telescope of the Square Kilometre Array Observatory (SKAO) \cite{Bacon:2018dui}. This multi-wavelength choice of four surveys is motivated by: a good coverage of redshifts in the range $0<z<2$; high redshift resolution in order to detect the Doppler effect; a negligible cross-shot noise between optical and 21cm intensity samples; and the very different systematics affecting optical and radio surveys, which are suppressed in cross-correlations.

We show that combining these surveys can provide a detection of the Doppler term with a signal-to-noise of $\sim$8. In the case of the lensing magnification contribution,
the H$\alpha$ survey on its own can deliver a 4\% error, while the full combination improves this to 2\%. For $\fnl$, the forecast multi-tracer constraint improves on Planck, at $\sigma(\fnl)\sim 1.5$, but falls short of the $\sigma(\fnl)\sim 1$ threshold. 
This is not unexpected, since our choice of surveys is based on the combination of relativistic and primordial non-Gaussian signals and is not optimised for $\fnl$ alone. {As an example, photometric} surveys or radio continuum surveys in a multi-tracer combination \cite{Alonso:2015sfa} {are more likely to achieve $\sigma(\fnl)< 1$, as they will have lower shot noise. Similarly,} combining the bispectrum and power spectrum of a single tracer  \cite{Karagiannis:2018jdt,Karagiannis:2019jjx,Karagiannis:2020dpq,Barreira:2021ueb}, {can achieve $\sigma(\fnl)< 1$, especially when using spectroscopic surveys \cite{Dore:2014cca,Tellarini:2016sgp,Karagiannis:2018jdt}. Furthermore, adding information from  higher-order point statistics, such as the trispectra of spectroscopic surveys \cite{Gualdi:2020eag}, will help reduce error bars on $\fnl$.}

We also find that the multi-tracer method approaches the maximum information that can be extracted, since the marginal errors approach the conditional errors.

The paper is organised as follows. In  \autoref{sec:threv} we review the theoretical aspects of the observed angular power spectrum, including scale-dependent bias and all relativistic effects.  \autoref{sec:modelsurvs} describes the survey specifications and astrophysical details. In   \autoref{sec:largescaleeffects} we focus on  local primordial non-Gaussianity, Doppler effects and  lensing magnification effects in a multi-tracer analysis of the observed angular power spectra. We present our results in  \autoref{sec:results} and conclude in  \autoref{sec:conclusion}. Throughout the paper we assume the fiducial cosmology: $A_s=2.142\times 10^{-9}$, $n_s=0.967$, $\odm=0.26$, $\ob=0.05$, $w=-1$, $H_0=67.74\,$ km/s/Mpc.

\section{The observed angular power spectra} \label{sec:threv}

We only consider scales where linear perturbation theory is valid.
The observed number of galaxies in a direction $\n$ within  solid angle $\d\Omega_{\n}$ and  redshift interval $\d z$ is 
\be
\d \mathbb{N}_{\rm g} =N_{\rm g}\, \d z\,\d\Omega_{\n}=  n_{\rm g}\, \d {\cal V} \,. 
\ee
Here  $N_{\rm g}$ is the   angular number density  per redshift as measured by the observer.  On the other hand, $n_{\rm g}$  is the comoving number density  and $\d {\cal V}$ is the comoving volume, which are not observed, since they are quantities at the emitting source. Then the observed number density contrast is given by \cite{Alonso:2015uua}
\be
\Delta_{\rm g}(z_i,\n,F_{\rm c})=\frac{N_{\rm g}(z_i,\n,F_{\rm c})-\bar{N}_{\rm g}(z_i,F_{\rm c})}{\bar{N}_{\rm g}(z_i,F_{\rm c})}\,,
\ee
where $F_{\rm c}$ is the survey flux cut.
The observed number density contrast is related to the number density contrast at the source, $\delta_{\rm g}=(n_{\rm g}-\bar n_{\rm g})/\bar n_{\rm g}$,  via redshift-space (light-cone) effects:
\bea \label{dobs}
 \Delta_{\rm g}  
=\delta_{\rm g} + \Delta^{\rm RSD} +\Delta^{\rm Dopp}  +\Delta^{\rm Lens}  +\Delta^{\rm Pot}\,.
\eea

The first term on the right of \eqref{dobs} is 
\be \label{nong}
\delta_{\rm g}(z,\n)=b(z)\,\delta_{\rm m}(z,\n)+2\fnl \, \delta_{\rm crit}\big[b(z)-1 \big]\varphi_{\rm p} (z,\n)\,,  
\ee
which is the number density contrast in comoving gauge -- i.e., it is defined physically, in the common rest frame of galaxies and dark matter  
\cite{Challinor:2011bk,Bruni:2011ta,Jeong:2011as,Alonso:2015uua}. Note that \eqref{nong} is a simplistic model that should be improved, as shown in \cite{Barreira:2020kvh,Barreira:2020ekm,Barreira:2021ueb}. For Gaussian primordial initial conditions ($\fnl=0$), it is proportional to the matter density contrast $\delta_{\rm m}$, where $b$ is the Gaussian clustering bias, which is scale-independent on the linear scales that we consider. The scale-dependent correction  from local primordial non-Gaussianity ($\fnl\neq0$) is given by the second term on the right of \eqref{nong}, where $\varphi_{\rm p}$ is the primordial Gaussian potential and $\delta_{\rm crit}=1.686$ is the threshold density contrast for spherical collapse. The primordial potential is related to the matter density contrast by linear evolution through the radiation and matter eras (see e.g. \cite{Maartens:2020jzf} for details). This is easier to express in Fourier space: 
\be\label{eq:bng}
\delta_{\rm g}(z,\k)=\left[b(z)+3 {\fnl}\, \frac{ \delta_{\rm crit}\big[b(z)-1 \big]{D(z_{\rm d})(1+z_{\rm d})\,\Omega_{\rm m0} H_0^2} }{D(z)\, T(k) \, k^2}\right]\delta_{\rm m}(z,\k)\,,
\ee
where  $T$ is the matter transfer function (normalised to 1 on ultra-large scales),  $D$ is the growth factor, normalised to 1 at $z=0$, and $z_{\rm d}$ is the redshift at decoupling (this is the CMB convention for $\fnl$).

In \eqref{dobs}, `RSD' denotes the linear (Kaiser)  redshift-space distortion:
\be
 \Delta^{\rm RSD}(z,\n) =-{1\over {\cal H}(z)}\n \cdot \nabla \big[\n \cdot \ve(z,\n)  \big] \,,
\ee
where ${\cal H}$ is the conformal Hubble rate and $\ve$ is the peculiar velocity.

The Doppler term is also sourced by peculiar velocity. It affects not only  the measured redshift but also has a (de-)magnification effect on galaxies \cite{Bonvin:2016dze}:
\bea\label{eq:DeltA_D}
 \Delta^{\rm Dopp}(z,\n,F_{\rm c}) &=& A_{\rm D}(z,F_{\rm c})\,\n \cdot \ve(z,\n) \,, \\ A_{\rm D}(z,F_{\rm c}) &=&\frac{\big[5s(z,F_{\rm c})-2\big]}{{\cal H}(z) \chi(z)}-5s(z,F_{\rm c})+b_{\rm e}(z,F_{\rm c})+\frac{\d \ln {\cal H}(z)}{\d \ln(1+z)},
\eea
where $\chi$ is the comoving radial distance.
The magnification bias $s$ is given by
\be
s(z,F_{\rm c})=-{2\over5}\frac{\partial \ln\bar N_{\rm g}(z,F_{\rm c})}{\partial \ln F_{\rm c}}= \frac{\partial \log\bar N_{\rm g}(z,m_{\rm c})}{\partial m_{\rm c}}\,,
\ee
where the second equality is given in terms of apparent magnitude $m$.
The evolution bias $b_{\rm e}$ quantifies how much the comoving number density deviates from constancy (e.g. due to mergers):
\be \label{bevo}
b_{\rm e}(z,F_{\rm c})=-\frac{\partial \ln \bar n_{\rm g}(z,L_{\rm c})}{\partial \ln (1+z)}\,,\quad
\bar n_{\rm g} = {(1+z){\cal H} \over \chi^2}\, \bar N_{\rm g}\,.
\ee
Here $L_{\rm c}$ is the luminosity threshold at the source, related to $F_{\rm c}$ via the luminosity distance: $L_{\rm c}=4\pi\,\bar{d}_L^{\,2}\,F_{\rm c}$. It is often convenient to use the total redshift derivative of number density, in which case we have the alternative expression for evolution bias \cite{Maartens:2021dqy}:
\bea\label{be2}
b_{\rm e}
=-{\d \ln \bar n_{\rm g}\over \d \ln (1+z)}- 5\left[1+{1 \over {\cal H}\chi}+ {2\ln 10\over 5}\,{\d K \over \d \ln(1+z)} \right]s\,,
\eea
where the $K$ is the K-correction, which vanishes for emission line surveys.

The `Lens' term in \eqref{dobs} denotes the large-scale lensing convergence contribution to the number density contrast:
\bea\label{eq:DeltA_L}
 \Delta^{\rm Lens}(z,\n,F_{\rm c}) &=&
 A_{\rm L}(z,F_{\rm c})\,\kappa(z,\n) \notag\\
&=& \big[5s(z,F_{\rm c}) -2\big]\,\int^{\chi(z)}_0{\rm d}\tilde\chi\,\frac{\big[\chi(z)-\tilde \chi\big]}{\chi(z)\,\tilde \chi}\, \nabla^2_{\n}{\Phi(\tilde \chi,\n)}\,,
\eea
where $\Phi$ is the Bardeen potential (in the standard cosmology, the metric potentials are equal), and $\nabla^2_{\n}$ is the 2-sphere Laplacian.

Finally, the `Pot' term  in \eqref{dobs} collects the remaining, sub-dominant relativistic potential effects: 
\bea
 \Delta^{\rm Pot}(z,\n,F_{\rm c}) &=& [A_{\rm L}(z,F_{\rm c})-A_{\rm D}(z,F_{\rm c})]\, \Phi(z,\n)+
 \frac{\partial \Phi( z,\n)}{\partial\ln(1+z)}
 +\big[3-b_{\rm e}(z,F_{\rm c})\big] {\cal H}V
 \nonumber\\\label{dpot}
&& -2A_{\rm L}(z,F_{\rm c})\int^{\chi(z)}_0 \! \frac{{\rm d}\tilde\chi}{\chi(z)}\,\Phi(\tilde \chi,\n)
  -2A_{\rm D}(z,F_{\rm c})
 \int^{\chi(z)}_0 \! {\rm d}\tilde\chi\,{\partial \Phi(\tilde \chi,\n) \over \partial\eta},
\eea
where $\eta$ is conformal time. The first two terms on the right  of \eqref{dpot} are Sachs-Wolfe terms. The third term is the Newtonian gauge correction that is needed for the physical definition of clustering bias, where $V$ is the velocity potential ($\partial_iV=v_i$). In line 2 of \eqref{dpot}, the first term is Shapiro time delay and the second is integrated Sachs-Wolfe.

In the case of 21cm intensity mapping, the observed HI brightness temperature  is related to the number of 21cm emitters per redshift per solid angle via \cite{Hall:2012wd,Alonso:2015uua}:
\be
T_{\HI}(z,\n)= {\rm const}\, { N_{\HI}(z,\n)  \over d_A^{\,2}(z,\n)}\,,
\label{tobs}
\ee
where $d_A$ is the angular diameter distance. This relation implies conservation of surface brightness, so that $\Delta_T$ has no lensing magnification contribution and $A_{\rm L}=0$. The evolution bias in terms of brightness temperature follows from \eqref{tobs} and \eqref{bevo}. Therefore 
the effective magnification bias and evolution bias for intensity mapping are
\bea
s &=& \frac{2}{5}\,, \label{tmag}\\
b_{\rm e} &=& -\frac{\d \ln \left[(1+z)^{-1}\H\bar{T}_{\HI} \right]}{\d \ln(1+z)}\,. \label{tevo} 
\eea

We use the angular power spectrum $C_\ell(z_i,z_j)$ as our summary statistic since it includes all relativistic  and wide-angle effects in a direct and simple way, as well as incorporating all cross-bin correlations. The angular power spectrum is related to the redshift-space 2-point correlation function via
\be\label{eq:Cldef}
\big\langle \Delta_{\rm g}(z_i,\n_i)\, \Delta_{\rm g}(z_j,\n_j)\big\rangle = \sum_{\ell}\,{(2\ell+1) \over 4\pi}\, C_\ell(z_i,z_j)\,{\cal L}_\ell \big(\n_i\cdot\n_j \big)\,,
\ee
where ${\cal L}_\ell$ are the Legendre polynomials. Expanding in spherical harmonics, we have 
\be
\Delta_{\rm g}(z,\n)= \sum_{\ell m}  \alm(z)\, \ylm{\n}\,,
\ee
where the covariance of the $\alm$ is the angular power spectrum
\be
\big\langle \alm(z_i) a^*_{\ell'm'}(z_j) \big\rangle=C_\ell(z_i,z_j)\ \delta_{\ell\ell'}\,\delta_{mm'}\,.
\ee
Then we find that
\be \label{eq:clgeneral}
C_\ell ( z_i,z_j )=4\pi\int{\rm d} \ln k\, \Delta_\ell( z_i,k)\, \Delta_\ell( z_j,k)\, \mathcal P (k)\,,
\ee
where $ \mathcal P (k)$ is the dimensionless primordial power spectrum and $\Delta_\ell$ are windowed transfer functions that relate $\Delta_{\rm g}$ to $ \mathcal P$.  
{The explicit relationship between the  redshift-space density contrast $\Delta_{\rm g}(z,\n)$ and the windowed transfer function $\Delta_\ell(z,k)$ can be found in \cite{DiDio:2013bqa,Alonso:2015uua}.}

When we consider multiple tracers $A=1,2,\cdots$, we define the generic angular power spectrum:
\bea\label{mtc1}
\big\langle \Delta_A(z_i,\n_i)\, \Delta_B(z_j,\n_j)\big\rangle &=& \sum_{\ell}\,{(2\ell+1) \over 4\pi}\, C^{AB}_\ell(z_i,z_j)\,{\cal L}_\ell \big(\n_i\cdot\n_j \big)\,, \\
C^{AB}_\ell ( z_i,z_j )&=& 4\pi\int{\rm d} \ln k\, \Delta_{A\ell}( z_i,k)\, \Delta_{B\ell}( z_j,k)\, \mathcal P (k)\,. \label{mtc2}
\eea

\section{Modeling next-generation cosmological surveys} \label{sec:modelsurvs}

We consider two galaxy surveys, with H$\alpha$ galaxy (high redshift) and bright galaxy (low redshift) samples, similar to Euclid and DESI surveys respectively (see \cite{2016arXiv161100036D,Euclid:2019clj} for further details). In addition, we consider the SKAO intensity mapping (IM) surveys in Bands 1 (high redshift) and 2 (low redshift), using the single-dish (autocorrelation) mode \cite{,Bacon:2018dui}. These surveys detect the 21cm emission line of neutral hydrogen (HI). They do not detect individual HI-emitting galaxies, but measure the integrated emission in each 3-dimensional pixel (formed by the telescope beam and the frequency channel width).

\begin{table}[!h]
\caption{\label{tab1} Volumes of next-generation surveys.}
\centering
\begin{tabular}{llcl}
\\ \hline\hline
Survey & Tracer &$\Omega_{\rm sky}$& Redshift \\
 & &$[10^3\deg^2]$ & range \\
\hline
{SKAO IM2} & {HI intensity map} & 20 & 0.10--0.58 \\
{SKAO IM1} & {HI intensity map} & 20 & 0.35--3.06 \\
Euclid-like & H$\alpha$ galaxies & 15 & 0.9--1.8 \\
DESI-like & Bright galaxies & 15 & 0.1--0.6 \\
\hline
\hline
\end{tabular}
\end{table}

In  \autoref{tab1} we summarise the footprint of the 4 surveys that we consider. The two SKAO surveys cover the same sky area. For the remaining cases, we assume that the overlap sky area for any pair is 10,000~deg$^2$.

\begin{figure}
 \centering
  \includegraphics[width=0.49\textwidth]{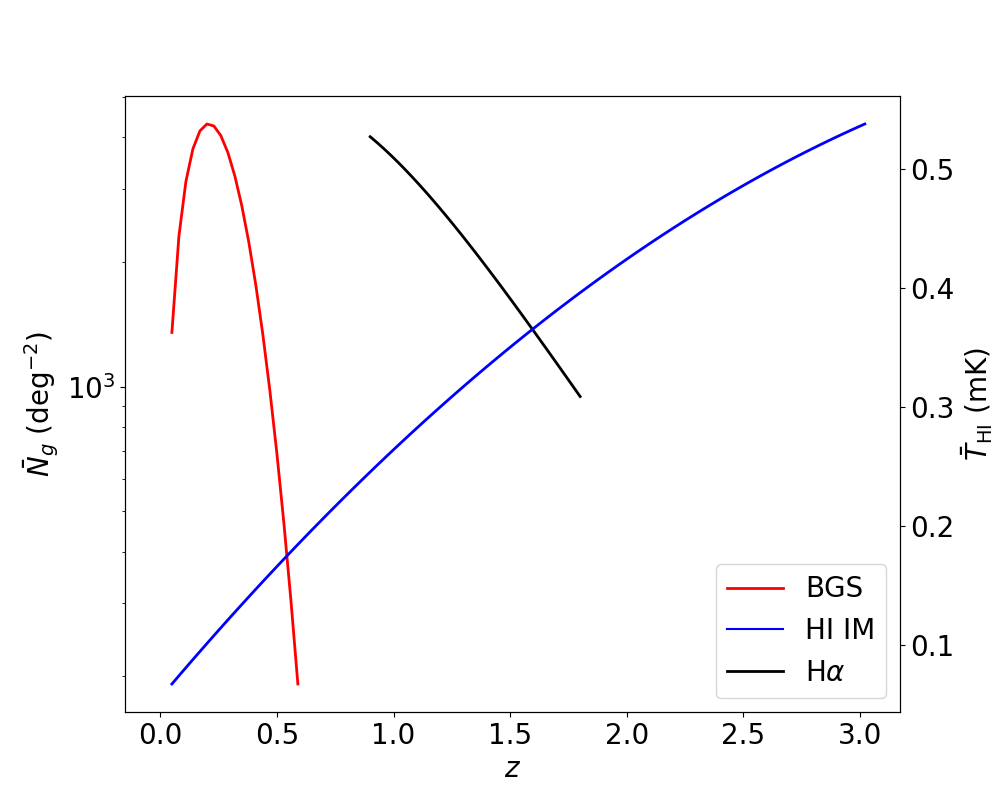}
  \includegraphics[width=0.49\textwidth]{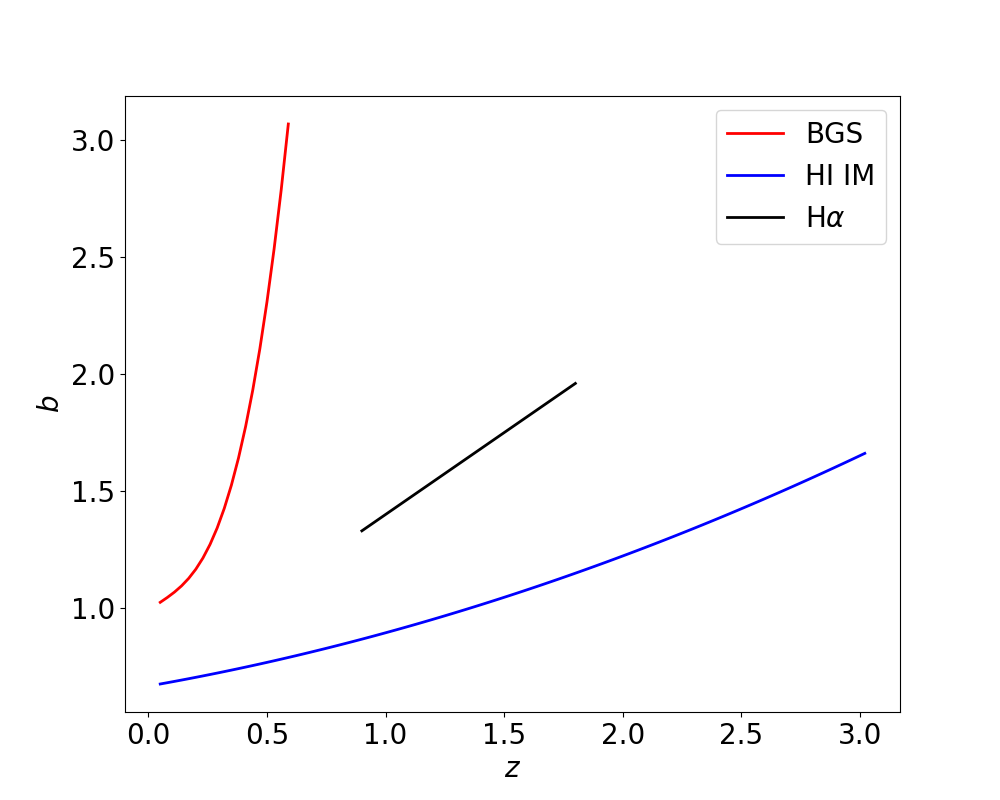}
  \includegraphics[width=0.49\textwidth]{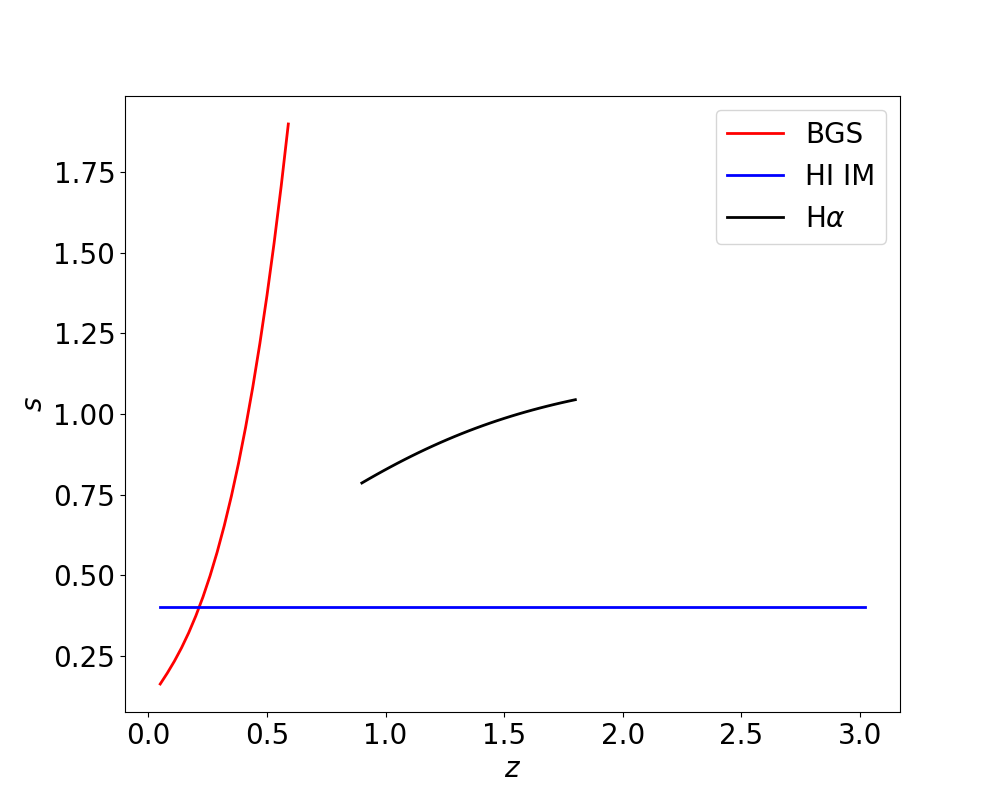}
  \includegraphics[width=0.49\textwidth]{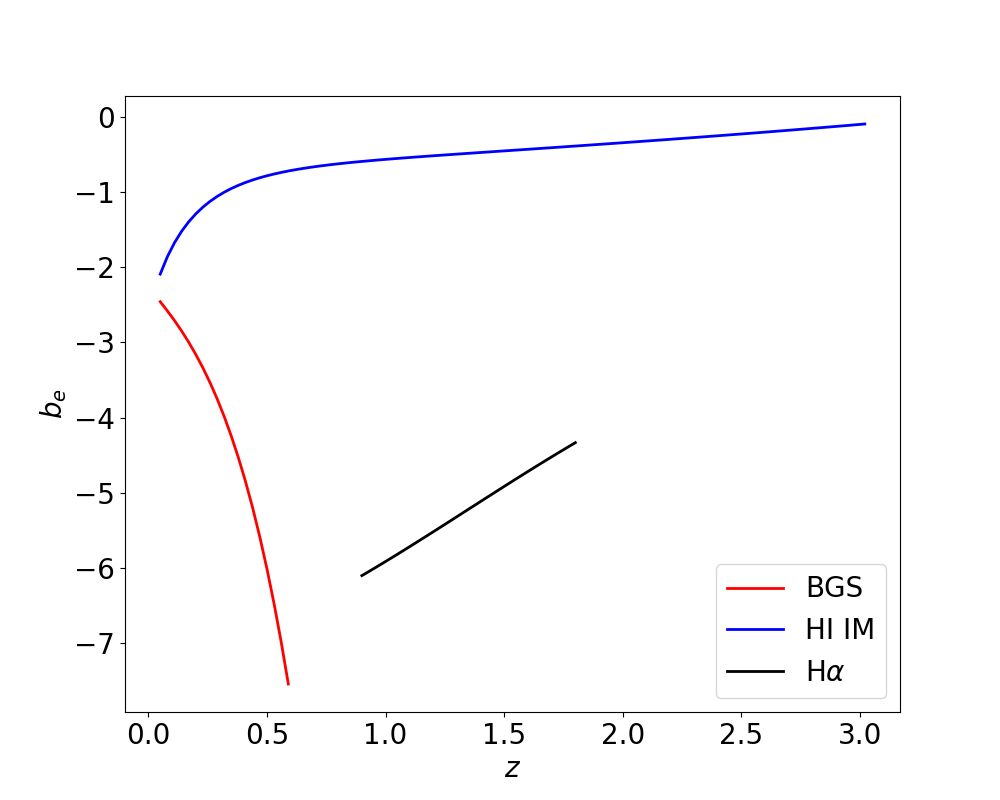}
  \caption{{\em Top:} Number density and HI temperature in mK ({\em left}), and Gaussian clustering bias ({\em right}), for the surveys. 
  {\em Bottom:} Magnification bias (\textit{left}) and evolution bias (\textit{right}), for the surveys.}\label{fig:sz_be}
\end{figure}

As described in \autoref{sec:threv}, to compute the angular power spectrum one requires astrophysical details such as 
number density, Gaussian clustering bias, magnification bias, evolution bias -- and the noise. {For Fisher forecasting of future galaxy surveys, we can use the appropriate model of the luminosity function, together with flux limits expected, to find $\bar n_{\rm g}$, $b_{\rm e}$ and $s$ in each redshift bin. Then fitting functions can be used to allow for different redshift binning. For IM surveys, $\bar n_{\rm HI}$, $b_{\rm e}$ and $s$ are survey-independent and we use a standard fitting function for the average brightness temperature $\bar T_{\rm HI}$ that is needed for $b_{\rm e}$. The details are given in \cite{Maartens:2021dqy}.} We used the $\bar n_{\rm g}$ in \cite{Maartens:2021dqy}  to determine fits for $\bar N_{\rm g}$ for the galaxy surveys. For the clustering biases, we use
\cite{2016arXiv161100036D} for BGS,  \cite{Merson:2019vfr} for H$\alpha$ and \cite{Bacon:2018dui} for  HI intensity mapping. 
The results are shown in \autoref{fig:sz_be} based on the following fits.
\begin{itemize}
\item
DESI-like BGS survey: 
\begin{eqnarray}
\label{eq:BGS_fit}
\bar N_{\rm g} &=& {z^{1.161} \exp\left[10.75-\left({}z \over 0.27\right)^{2.060}\right]} \,, \\
b&=& 0.99+0.73\,z-1.29\,z^2+10.21\,z^3 \,,\\
s &= & 0.113 + 0.945\,z + 0.908\,z^2 + 4.442\,z^3  \,,  \label{fqd}\\
b_{\rm e} &= &  -2.25   -4.02\,z +0.318\,z^2 -14.6\,z^3  \,.  \label{fbed}
 \label{fbld}
\end{eqnarray} 
\item
Euclid-like H$\alpha$ survey:
\begin{eqnarray}\label{eq:Ha_fit}
\bar N_{\rm g} &=& z^{1.985} \exp\left(0.019\,z^4-0.052\,z^3+0.147\,z^2-3.405\,z+11.471\right)\,,\\
b&=&0.7(1+z) \,, \label{fble} \\
s &=& 0.234 + 0.801\,z - 0.222\,z^{2} + 0.015\,z^{3}\,, \label{fqe}\\
b_{\rm e} &=& -7.29 + 0.470\,z + 1.17\,z^{2} - 0.290\,z^{3}\,. \label{fbee} 
\end{eqnarray}

\item
HI intensity mapping (any survey):
\bea
\bar{T}_{\HI} &=& 0.0559 + 0.2324\,z - 0.0241\,z^2~~{\rm mK} \,,\\
b &=& 0.667 + 0.178\, z + 0.050\, z^2\,, \\
s &=& 0.4 \,,\\
 b_{\rm e}  &=&  -0.32-0.11\,z + 0.06\,z^2 -\exp\left(-0.9\,z^3+3.12\,z^2-4.61\,z+0.78\right).
\eea
\autoref{fig:sz_be} shows that the evolution of  sources in  intensity mapping is less susceptible to galaxy mergers, as compared to the  number counts in galaxy surveys, since intensity mapping does not resolve individual galaxies. As a result, $|b_{\rm e}^{\rm IM}|<|b_{\rm e}^{g}|$.
\end{itemize}

The intensity mapping angular power spectrum is modulated by the telescope beam, which in single-dish mode leads to a loss of small-scale transverse power:
\bea
C^{\rm IM\,IM}_\ell(z_i,z_j)~~&\to&~~\beta_\ell(z_i)\, \beta_\ell(z_j)\,C^{\rm IM\,IM}_\ell(z_i,z_j) \,,\\
\beta_\ell(z_i) &=& \exp\left[-{\ell(\ell+1) \over 16\ln2}\theta_{\rm b}^2(z_i) \right],
\eea
where $\theta_{\rm b}=1.22 \lambda_{21} (1+z)/D_{\rm d}$ is the effective field of view and $D_{\rm d}$ is the dish diameter.

Finally, we model
the uncertainty due to noise, which limits how well the signal can be extracted. For galaxy surveys, the angular shot-noise power is inversely proportional to the number of galaxies observed per solid angle:
\be \label{eq:noiseGS}
\mathcal{N}^{\rm g}_{\ell}(z_i,z_j) =\frac{1}{N_\Omega(z_i)} \, \delta_{ij} ~~\mbox{where}~~ N_\Omega(z_i)=\int \d z\, W(z,z_i)\bar N_{\rm g} (z)\,.
\ee
The window function $W$ is {a top-hat like window smoothed on the edges, given by \cite{Fonseca:2019qek}
\be \label{eq:Wths}
W(z,z_i)=\frac1{2\,{\tanh}({\Delta} z_i/2\sigma_{zi})} \left[  \tanh\!\left( \!\frac{z-z_i+{\Delta} z_i/2}{\sigma_{zi}}\!\right)\! -\! \tanh\!\left( \!\frac{z-z_i-{\Delta} z_i/2}{\sigma_{zi}}\! \right)\! \right]\!,
\ee
where ${\Delta} z_i$ is the redshift bin size, $\sigma_z$ is the redshift resolution and $\sigma_{zi} =\sigma_z(1+z_i)$.} For HI IM experiments the main noise source is instrumental and we can neglect shot noise on the scales of interest.
For single-dish surveys, the noise {is \cite{Knox:1995dq,Bull:2014rha}}
\be\label{eq:noiseIM}
\mathcal{N_{\ell}^{\rm IM}}(z_i,z_j)=\frac{4\pi \, f_{\rm sky} \, T^2_{\rm sys}(z_i)}{2 N_\d\,  t_{\rm tot}\,\Delta \nu} \, \delta_{ij}\, .
\ee
The system temperature $T_{\rm sys}$ is frequency dependent, $N_\d$ is the number of dishes, the frequency band of observation is $\Delta\nu$, the total integration time is $t_{\rm tot}$, and the sky fraction is $f_{\rm sky}=\Omega_{\rm sky}/4\pi$. We fix the scanning ratio, i.e the sky area over time. This implies that the observational time $t_{\rm tot}$, needs to be adjusted proportionally to the reduction in sky area. Further details on subtleties in the SKAO noise are given in \cite{Fonseca:2019qek}. 

Given the overlap between the BGS and IM2 surveys, and between the H$\alpha$ and IM1 surveys, we need to consider cross-shot noise. However, as shown in  \cite{Viljoen:2020efi}, the cross-shot noise is negligible and may be neglected.

\begin{figure}
 \centering
\includegraphics[width=0.6\textwidth]{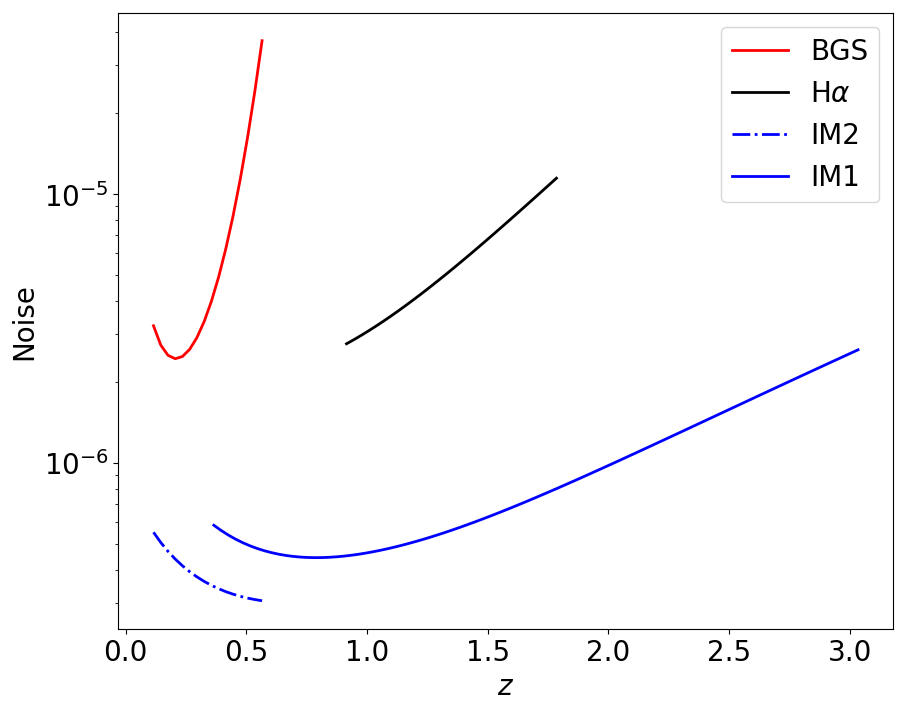}
  \caption{Shot noise  for galaxy surveys and instrumental noise (normalised by $\bar{T}_{\HI}^2$)  for intensity mapping surveys.}\label{fig:Noise}
\end{figure}

The noise power spectra for the various surveys are displayed in  \autoref{fig:Noise}. 

\section{Primordial non-Gaussianity and relativistic effects in the multi-tracer} \label{sec:largescaleeffects}

The scale-dependent clustering bias in Fourier space, given by \eqref{eq:bng}, grows on large scales as $\fnl H_0^2/k^2$, producing a power spectrum monopole $P_{\rm g}(z,k)$ that blows up on ultra-large scales. This behaviour in Fourier space is misleading, since the {directly} measurable signal of $\fnl$ is very small. The point here is that the Fourier power spectrum is not {directly} observable. {One has to de-project measured angular positions and redshifts into a Cartesian three-dimensional volume -- which requires assuming a fiducial cosmology. The Fourier power spectrum is a derived quantity, which is also not gauge invariant.} By contrast, the angular power spectrum is observable (and this is also why no Alcock-Paczynski correction is needed when analysing data via the angular power spectrum).

\begin{figure}
 \centering
  \includegraphics[width=0.49\textwidth]{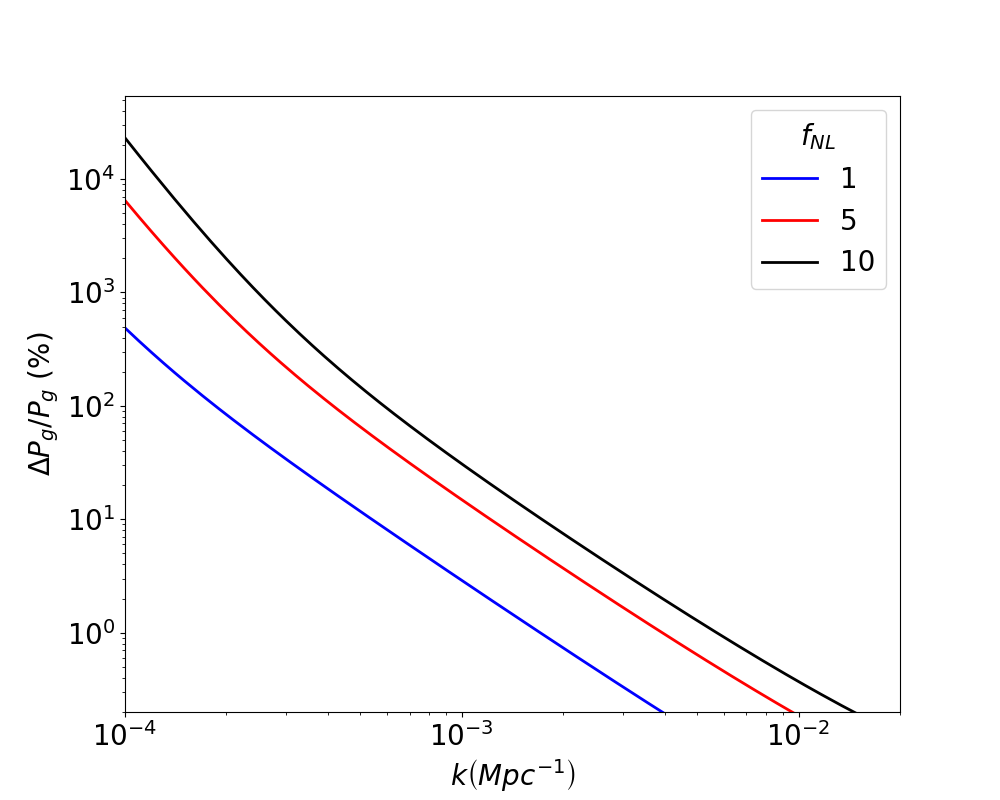}
  \includegraphics[width=0.49\textwidth]{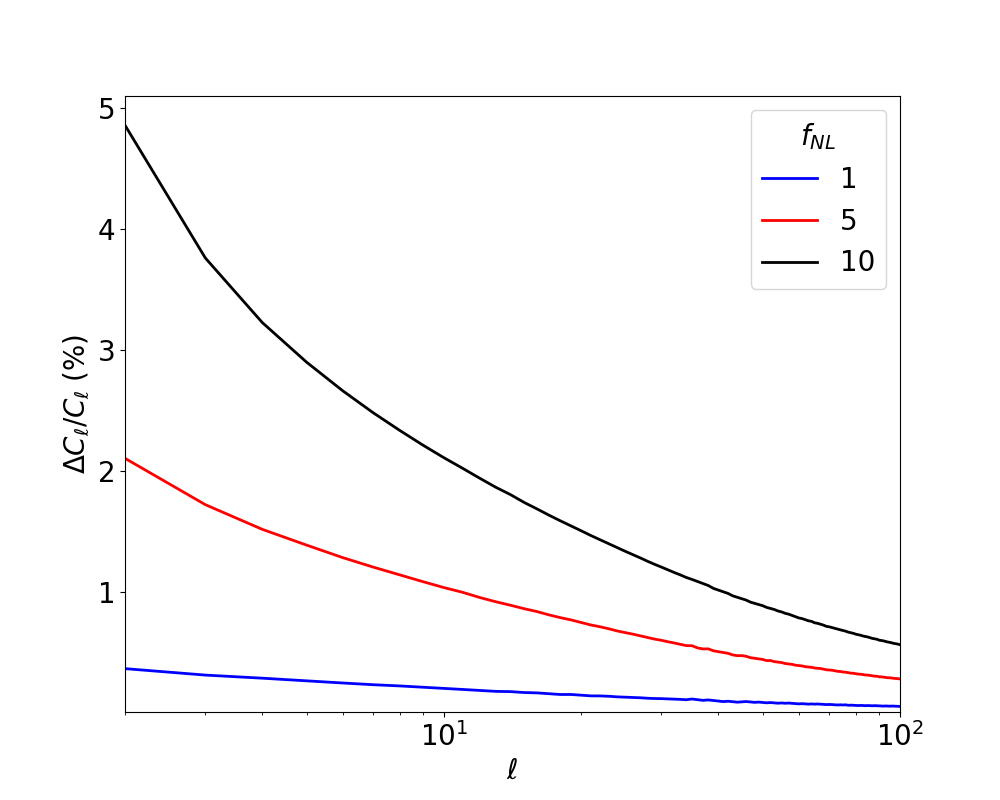}
  \caption{Percentage contribution of scale-dependent bias to the Fourier power spectrum monopole (\textit{left}) and angular power spectrum (\textit{right}) for an H$\alpha$ survey at $z=1.8$ with $\Delta z=0.03$. }\label{fig:Pk_Cl_fnl}
\end{figure}

We illustrate this point in \autoref{fig:Pk_Cl_fnl}, showing an example of the percentage contribution of $\fnl$ to the Fourier and angular power spectra.
The blow-up in the unobservable $P_{\rm g}$ is not mirrored by the behaviour of $C_\ell$, in which the $\fnl$ contribution is only a few percent on the largest scales, where cosmic variance is typically much larger. This is why we require a multi-tracer approach to detect the signal of $\fnl$.

\begin{figure}
 \centering
  \includegraphics[width=0.49\textwidth]{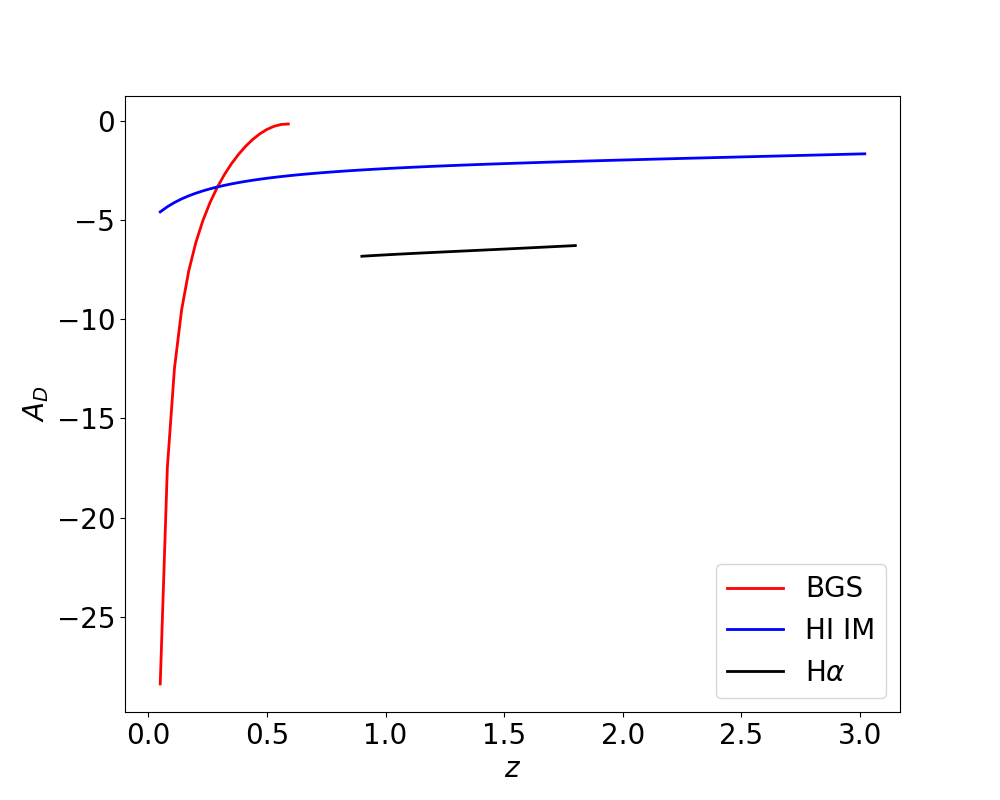}
  \includegraphics[width=0.49\textwidth]{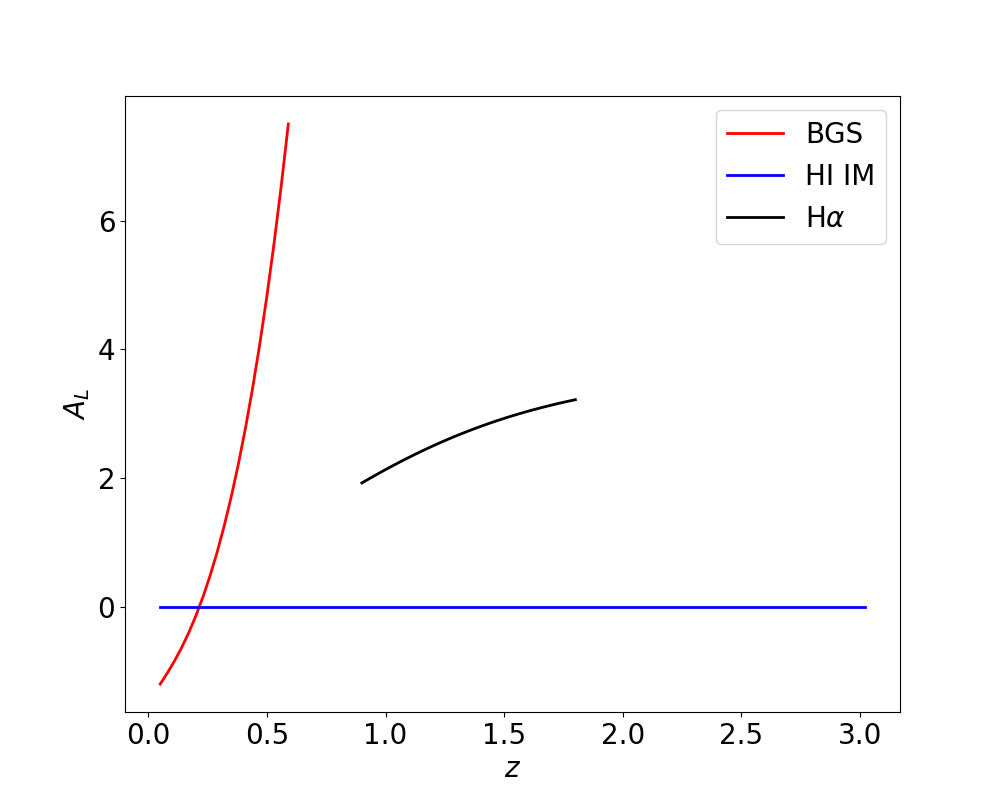}
  \caption{Coefficient of the Doppler ({\em left}) and lensing ({\em right}) contributions  for the surveys.}
  \label{fig:AD_all}
\end{figure}

The dominant relativistic effects are from the Doppler and lensing magnification terms. The Doppler term depends on both the magnification and evolution biases via \eqref{eq:DeltA_D}, while the lensing term \eqref{eq:DeltA_L} depends only on the magnification bias. Using the fits in  \autoref{sec:modelsurvs}, we show in  \autoref{fig:AD_all} the amplitude of the Doppler and lensing contributions to the observed number density and temperature contrasts. 

It is apparent that the Doppler coefficient for intensity mapping is significantly smaller than for the galaxy samples.  Therefore we expect HI surveys to under-perform in extracting information on the Doppler effect, compared to galaxy surveys. We also note that the Doppler amplitude is strongest at low redshift, so that the BGS survey is best placed to detect it.

The lensing coefficient vanishes for intensity mapping and is largest for the BGS survey. This does not mean that the lensing contribution for BGS is greater than for H$\alpha$, since the lensing amplitude is $|A_{\rm L}\kappa|$ and $\kappa$ will be larger at higher $z$.

For our Fisher analysis, we separate the relativistic contributions by introducing fudge factors $\varepsilon=1$ to gauge the detectability of each contribution:
\be \label{eq:defeps}
 \Delta_A =\delta_A + \Delta^{\rm RSD} + \varepsilon_{\rm D}\,\Delta^{\rm Dopp} + \varepsilon_{\rm L}\,\Delta^{\rm Lens} + \varepsilon_{\rm P}\,\Delta^{\rm Pot}\,.
\ee
In order to determine which correlations, both in single- and multi-tracer cases, 
add the most information, 
we compute the signal-to-noise of a parameter $\vartheta$, constrained by two surveys, 
from each correlation ${C}^{AB}_{\ell}(z^A_i,z^B_j)$, as
\be \label{eq:SAB}
S^{AB}(\vartheta,z^A_i,z^B_j) = \left\{ \sum_\ell \left[\frac{\partial_{\vartheta} \, {C}^{AB}_{\ell}(z^A_i,z^B_j)}{{\Delta C}^{AB}_{\ell}(z^A_i,z^B_j)}\right]^{2}\right\}^{1/2} \,,
\ee
where $A,B=1,2$ and
\be
{\Delta C}^{AB}_{\ell}(z^A_i,z^B_j) =\left\{\frac{\Gamma^{AA}_{\ell}(z^A_i,z^A_j)\Gamma^{BB}_{\ell}(z^B_i,z^B_j)+\left[\Gamma^{AB}_{\ell}(z^A_i,z^B_j)\right]^2}{(2\ell+1)f_{\rm sky}} \right\}^{1/2}\,,
\ee
with
\be\label{eq:matrix_Cl_Gamma}
\Gamma^{AB}_\ell =C^{AB}_\ell+{\mathcal N}^{AB}_\ell\,,
\ee
{where the noise is  given by (\ref{eq:noiseGS}) for galaxy surveys and (\ref{eq:noiseIM}) for IM.}

\begin{figure}
 \centering
\includegraphics[width=0.49\textwidth]
    {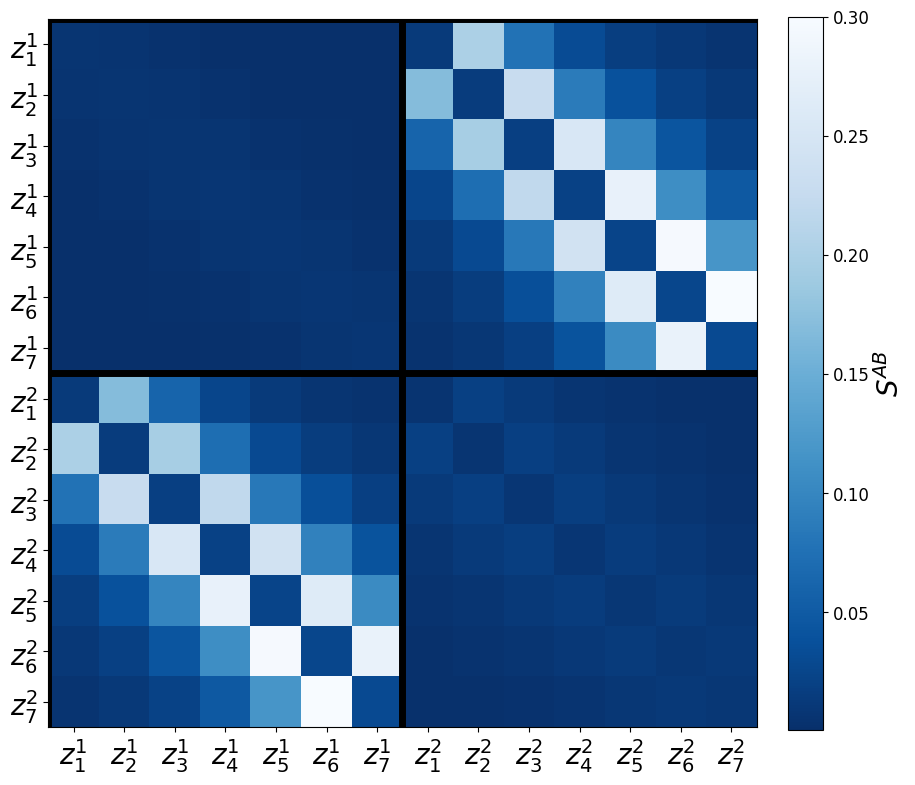}
\includegraphics[width=0.49\textwidth]
    {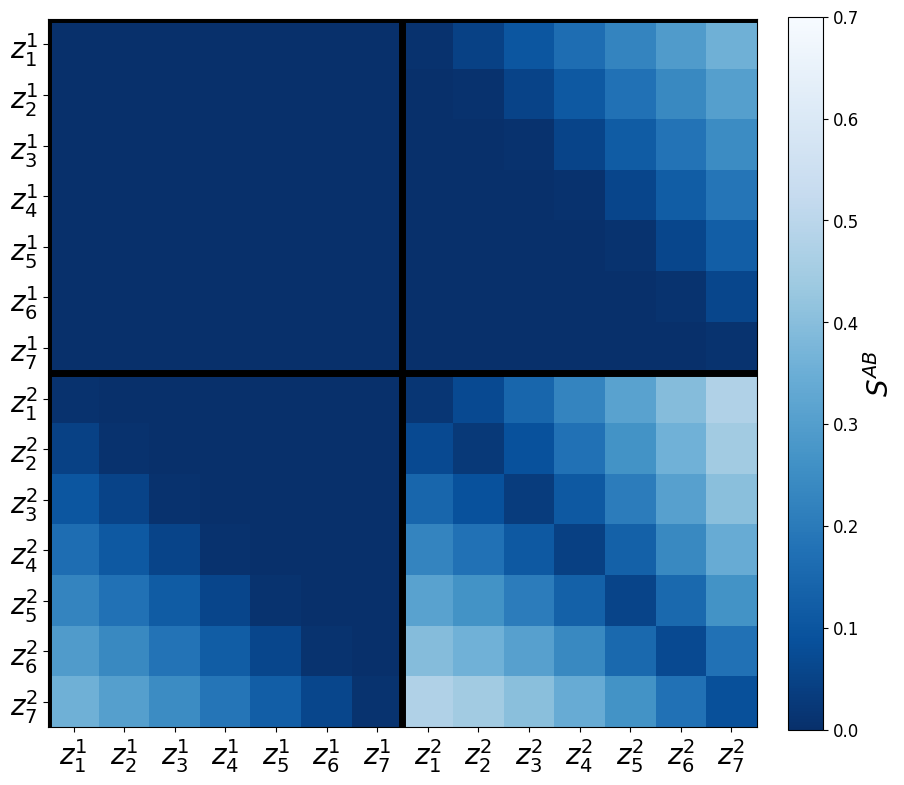}
\includegraphics[width=0.49\textwidth]
    {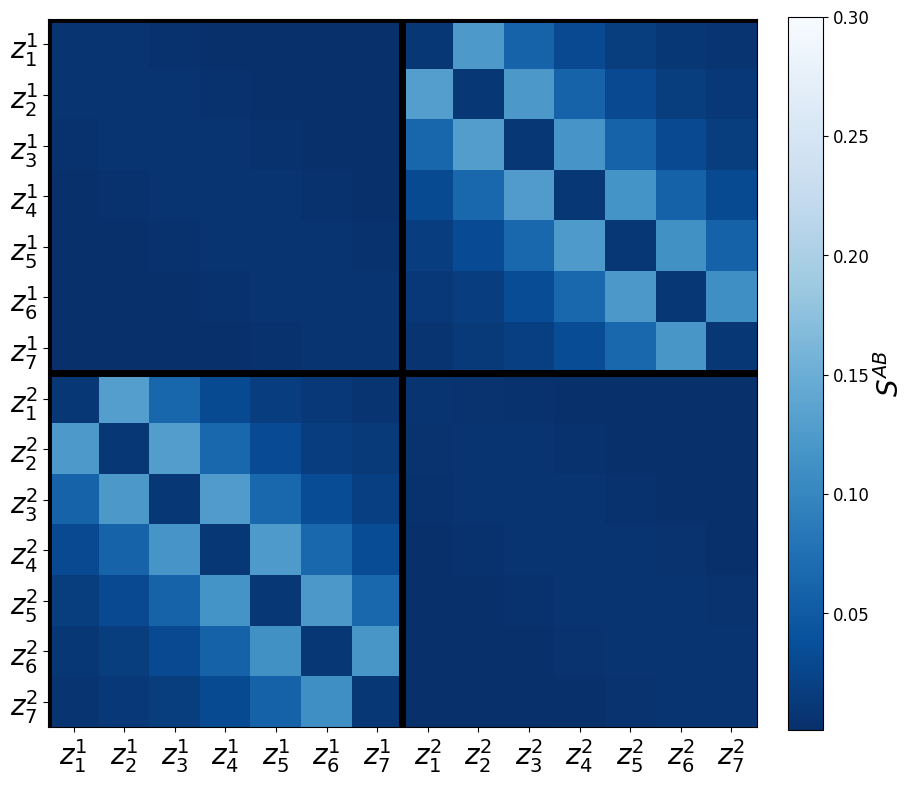}
\includegraphics[width=0.49\textwidth]
    {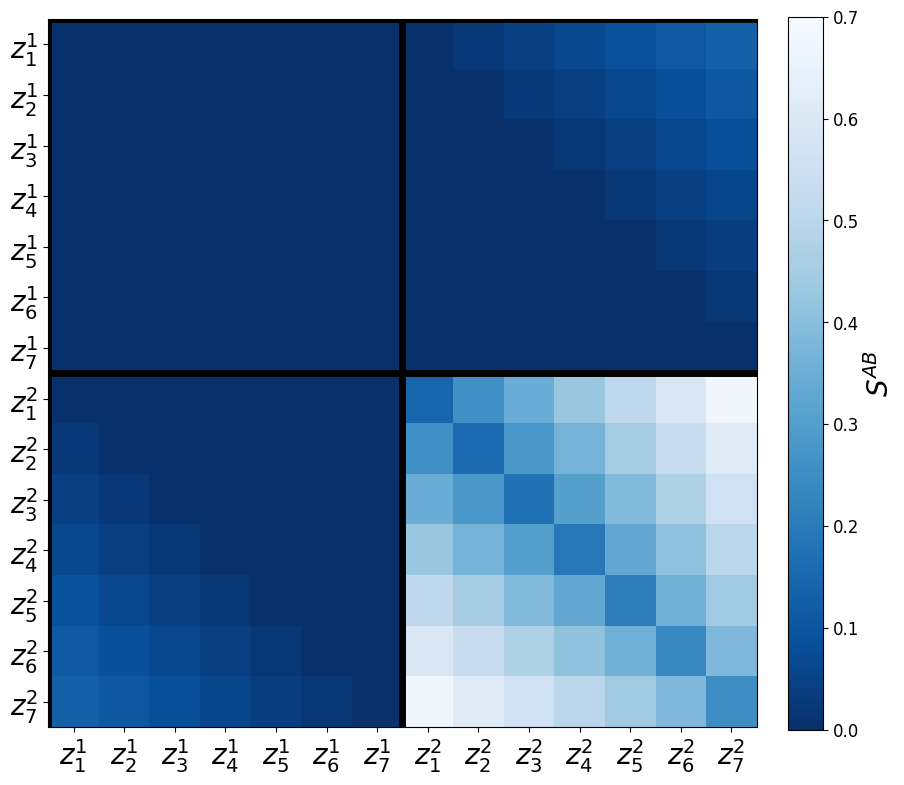}
  \caption{Signal-to-noise $S^{AB}(\theta,z^A_i,z^B_j)$ for the Doppler contribution ($\theta=\varepsilon_{\rm D}$) (\textit{left}) and the lensing magnification contribution ($\theta=\varepsilon_{\rm L}$) (\textit{right}). \textit{Top panels:} $1\otimes 2= {\rm IM2}\otimes {\rm BGS}$ for $0.35< z<0.56$. \textit{Bottom panels:} $1\otimes 2= {\rm IM1}\otimes {\rm H}\alpha$ for $0.90<z<1.15$. Colour bar shows  the signal-to-noise.}\label{fig:SN_D_fnl}
\end{figure}

As we confirm below, the potential terms are very poorly constrained, and we set $\varepsilon_{\rm P}=0$ in this discussion.
In \autoref{fig:SN_D_fnl} we show examples of $S^{AB}$ for $\vartheta=\varepsilon_{\rm D}$ ({left}) and $\vartheta=\varepsilon_{\rm L}$ ({right}) using both low-redshift ({top}) and high-redshift  ({bottom}) survey pairs. The axes indicate which $z$-bins are correlated, where $z^1_i$ is the $i$-bin of SKAO IM2 (top panels) and IM1 (bottom panels), while $z^2_i$ corresponds to BGS (top panels) and H$\alpha$  (bottom panels). For the multi-tracer combination  $1\otimes2 ={\rm IM2}\otimes {\rm BGS}$ (top panels) we use the redshift range $0.35<z<0.41$, and for $1\otimes 2={\rm IM1}\otimes {\rm H}\alpha$ ({bottom panels}), we use $0.9<z<1.12$. In both cases a bin-width $\Delta z=0.03$ is used. 
The colour bar shows the signal-to-noise $S^{AB}$ of the parameter for each pair $(z^A_i,z^B_j)$.

The plots show that the bulk of the information on $\varepsilon_{\rm D}$ comes from the cross-spectra, which contribute 3--5 times more than the individual surveys. This contribution is mainly from the cross-tracer spectra, although at low redshift it is also marginally from cross-correlations between consecutive redshift bins of the same tracer. The latter feature is the basis of the estimator constructed in \cite{Fonseca:2021gve}, which is the angular power spectrum counterpart of the Doppler dipole in the 2-point correlation function \cite{Bonvin:2013ogt}. Although there is some residual information in the low-redshift BGS-BGS correlations, the bulk clearly comes from the off-diagonal, i.e. the cross-tracer correlations. 
BGS has most information on the Doppler term,  given the $1/\chi$ term in \eqref{eq:DeltA_D} (see \autoref{fig:AD_all}). 

In the case of lensing magnification, the analysis is slightly different. Firstly,  IM has no lensing magnification contribution ($A_{\rm L}=0$). However, IM can act as a lens of background galaxies in BGS and H$\alpha$. In both cases, we see that the furthest off-diagonal cross-bin correlations are the ones that provide most information on $\varepsilon_{\rm L}$. As expected this comes from the lower redshift field lensing the higher redshift field. For this reason, when the higher redshift field is traced by IM, no information is obtained -- as seen in the right column of  \autoref{fig:SN_D_fnl}. We also see that most information on $\varepsilon_{\rm L}$ is obtained by galaxy surveys on their own, especially at high redshift. In fact, for the H$\alpha$ survey,  cross-correlations with an IM survey seem to be irrelevant for lensing. Most information lives in the furthest off-diagonal cross-bin correlations of H$\alpha$-H$\alpha$.

Having gained qualitative insights into  the multi-tracer
precision on $\fnl$ and relativistic effects, we turn now to the quantitative
estimate of constraints via Fisher forecasting.
We will also include constraints on the standard cosmological parameters.

The multi-tracer Fisher matrix for angular power spectra is defined as follows \cite{Alonso:2015sfa,Fonseca:2015laa,Viljoen:2020efi}:
\bea \label{eq:fishercl}
F_{{\vartheta}_{\alpha}{\vartheta}_{\beta}}=\sum_{\ell_{\rm min}}^{\ell_{\rm max}} \frac{(2\ell+1)}{2 }f_{\rm sky}\,{\rm Tr}\Big[ \big(\partial_{{\vartheta}_{\alpha}} \bm{C}_{\ell}\big)\, \bm{\Gamma}_\ell^{-1} \big(\partial_{{\vartheta}_{\beta}}\bm{C}_{\ell}\big)\,\bm{\Gamma}_\ell^{-1}\Big]\,,
\eea
where $\vartheta_\alpha$ are the parameters and
\be\label{eq:matrix_Cl_Gamma}
\bm{\Gamma}_\ell 
=\bm{C}_\ell+\bm{\mathcal N}_\ell\,.
\ee
For $N$ tracers: 
\bea
\bm{ C}_\ell
=
\begin{bmatrix}
  { C}^{11}_\ell ( z^1_{i},  z^1_{j}) &  ~~~{C}^{12}_\ell ( z^1_{i},  z^2_{j}) & ~~~\cdots & ~~~{C}^{1N}_\ell ( z^1_{i},  z^N_{j})\\&&& \\~~~\cdots & ~~~\cdots&~~~\cdots &~~~\cdots \\ &&&\\
  {C}^{N1}_\ell ( z^N_i,  z^1_j) &  ~~~{C}^{N2}_\ell ( z^N_i,  z^2_j)& ~~~\cdots
  & ~~~{C}^{NN}_\ell ( z^N_{i},  z^N_{j})
\end{bmatrix}\,, \label{eq:cov_overlap}
\eea
{while the noise is the diagonal matrix
\bea
\bm{ \mathcal N}_\ell
=
\begin{bmatrix}
  { \mathcal N}^{11}_\ell ( z^1_{i},  z^1_{j}) &  ~~~0 & ~~~\cdots & ~~~0\\&&& \\~~~\cdots & ~~~\cdots&~~~\cdots &~~~\cdots \\ &&&\\
  0 &  ~~~0& ~~~\cdots
  & ~~~{\mathcal N}^{NN}_\ell ( z^N_{i},  z^N_{j})
\end{bmatrix}\,, \label{eq:noise_matrix}
\eea
where we use (\ref{eq:noiseGS}) for galaxy surveys and (\ref{eq:noiseIM}) for intensity mapping. In the Fisher matrix summation we take $\ell_{\rm min}=5$, while $\ell_{\rm max}$ is redshift dependent. We follow \cite{Fonseca:2019qek} and impose
\be
\ell_{\rm max}(z_i) = \chi(z_i) k_{\rm nl}(z_i)\,,
\ee
where $k_{\rm nl}(z)=0.2 (1+z)^{2/(2+n_s)}~ h\,{\rm Mpc}^{-1}$ \citep{Smith:2002dz}, the scale marking the breakdown of a perturbative analysis of matter clustering.}

In our forecasts we consider the set of parameters
\be \label{pars}
{\vartheta}_{\alpha}=\Big\{ \fnl,\, \varepsilon_{\rm D},\, \varepsilon_{\rm L},\, \varepsilon_{\rm P};\, \Omega_{\rm cdm},\, \Omega_{\rm b},\, w,\, n_s,\, H_0,\, A_s; \, {b_A(z_i^A)} \Big\}\,.
\ee
Note that the $\varepsilon_I$ do not have physical meaning; instead they are used to model the detectability of the relativistic contributions. 
{We also consider the uncertainty in the Gaussian clustering biases by marginalising over the bias in each redshift bin, $b_A(z_i^A)$}. In the single-tracer case, the forecasts are 
{degraded by this marginalisation, but} 
the multi-tracer analysis is only weakly affected by marginalisation over the clustering bias, consistent with \cite{Fonseca:2015laa}. We do not marginalise over the magnification and evolution biases. Instead, we consider  $\varepsilon_{\rm L}$ and $\varepsilon_{\rm D}$  as rough proxies of the uncertainties in these astrophysical parameters. 

\section{Results} \label{sec:results}

\begin{table}[!ht]
\caption{Marginal uncertainties on $\fnl$ and relativistic effects $\varepsilon_I$,  computed by marginalising over the standard cosmological and $b_A(z_i)$ only. Results are for individual surveys and their combination using the multi-tracer technique ($\otimes$). When combining low- and high-redshift information, we also consider adding their Fisher information matrices ($\oplus$).  HI intensity mapping surveys are unaffected by lensing and cannot constrain $\varepsilon_{\rm L}$. (Results exclude priors.) }\label{tab:err_mar_CP}
\centerline
{
\begin{tabular}{clccccc}
\\ \hline\hline
Redshift & Survey & {$\sigma(\fnl)$} &  {$\sigma(\varepsilon_{\rm D})$}  & {$\sigma(\varepsilon_{\rm L})$} & $\sigma(\varepsilon_{\rm P})$\\
\hline
$0.1-0.58$ & {BGS} & 26.38  & 7.57  & 0.39  & 33.3 \\
& {IM2}  & 35.74  & 18.07  & $-$  & 228.3 \\
& {IM2$\otimes$BGS}  & 2.12  & 0.14  & 0.13  & 6.86 \\
\hline
$0.9-1.8$ & {H$\alpha$}  & 9.34  & 9.08  & 0.04  & 10.0\\ 
$0.35-3.05$ & {IM1} & 4.72  & 6.29  & $-$  & 10.89 \\
$0.60-3.05$ & {IM1$\otimes$H$\alpha$} & 3.06  & 0.37  & 0.03  & 4.71\\
\hline
$0.1-3.05$ & {(IM2$\otimes$BGS)$\oplus$(IM1$\otimes$H$\alpha$)} & 1.70  & 0.13  & 0.03  & 3.79\\
& {IM2$\otimes$BGS$\otimes$IM1$\otimes$H$\alpha$} & 1.55  & 0.13  & 0.02  & 3.47\\
\hline\hline
\end{tabular}
}
\end{table}

The {forecast} precision on $\fnl$ and relativistic effects from future surveys is computed for individual surveys, as well as {for the multi-tracer case} that includes all correlations between different surveys.
In  \autoref{tab:err_mar_CP} we show the marginal error on $\fnl$ and $\varepsilon_I$, marginalised over
the uncertainty of the standard cosmological parameters and the biases $b_A(z_i)$. This approach is in some sense closer to reality, since including the light-cone corrections in the transfer function is the correct model to use. The  caveat is that this assumes the magnification and evolution biases are known. Later we will partly incorporate uncertainty in these biases by including marginalisation over the  $\varepsilon_I$, which can be seen as a `marginalisation' over the amplitude uncertainties of the relativistic terms. 

The highest precision on $\fnl$ from an individual survey is {$\sigma(\fnl)=4.72$}  from SKAO IM1. This is not surprising since IM1 boasts the largest observed volume among the surveys. The bigger the volume, the larger the scales we observe and  the more frequently we can sample correlations at scales that carry an $\fnl$ signal (see \autoref{fig:Pk_Cl_fnl}). 
The IM1 precision on $\fnl$ can be improved by a factor of 3 by combining all low- and high-$z$ surveys  (IM2$\otimes$BGS$\otimes$IM1$\otimes$H$\alpha$), resulting in  {$\sigma(\fnl)=1.55$}. This is significantly better than current constraints. 

It is interesting to note that individually the high-$z$ surveys constrain $\fnl$ remarkably  better than the low-$z$, since the volumes observed are considerably larger.
Individually the low-$z$ surveys cannot access as many correlations on large scales and therefore cannot constrain $\fnl$ well. {However, the multi-tracer eliminates cosmic variance, which enables the low-$z$ combination to access ultra-large scales and extract more $\fnl$ signal}. This is a clear example of how the multi-tracer works. By cancelling cosmic variance, dark matter tracer combinations in smaller overlapping volumes can perform as well as, or better than, single tracers covering large volumes. In fact the low-$z$ combination IM2$\otimes$BGS does better than its IM1$\otimes$H$\alpha$ counterpart, because of the reduced noise at lower redshift, as shown in  \autoref{fig:Noise}.

The best single-survey constraint on the Doppler contribution is again from the SKAO IM1 survey, with  {$\sigma(\varepsilon_{\rm D})=6.29$}. 
This survey has the largest volume, which compensates for the small amplitude of its Doppler term \eqref{eq:DeltA_D}. The BGS survey has the highest amplitude: its Doppler coefficient is largest (see \autoref{fig:AD_all}), and the radial peculiar velocity $\bm n \cdot \bm v$ is largest at low $z$. Nevertheless, the volume beats the amplitude for a term that has signal on ultra-large scales. 
As soon as we multi-trace the surveys, again we see the domination of survey volume fades away, due to cancellation of cosmic variance. Similar to $\fnl$, the low-$z$ combination outperforms the high-$z$ combination by a factor of nearly 3, because of the lower noise at low $z$.
This effect can also be seen in  \autoref{fig:SN_D_fnl}. The full combination of four surveys offers a precision of $\sigma(\varepsilon_{\rm D})=0.13$, only slightly better than the low-$z$ combination. The multi-tracer improvement in precision over the best single tracer is a factor of nearly {50}. This precision on the Doppler amplitude corresponds to a signal-to-noise of $\sim$8, sufficient for a detection of the Doppler term.
It also suggests that we can make some constraint on the evolution  bias, assuming that the  magnification bias is constrained by the lensing term (see below).

HI intensity mapping 
is unaffected by lensing magnification and so cannot constrain the lensing term. The best single-survey constraint is from the high-redshift H$\alpha$ survey, as expected, delivering $\sigma(\varepsilon_{\rm L})=0.04$. This precision is about double that found in \cite{Cardona:2016qxn}. The reason is that we use many more redshift bins in our Fisher analyses than they use in their MCMC.
Even though IM does not constrain lensing by itself, its correlation with a galaxy survey does improve the error~-- since 
galaxies at $z_i$ {behind IM at $z_j<z_i$ are lensed by the IM}.
This is confirmed by \autoref{fig:SN_D_fnl}. As a result, IM2$\otimes$BGS improves the BGS-only lensing constraint by a factor more than 2, while IM2$\otimes$H$\alpha$ improves on H$\alpha$-only by $\sim$25\%. The correlation of all four surveys improves on the H$\alpha$-only constraint by a factor of 2, with $\sigma(\varepsilon_{\rm L})=0.02$. This is a high enough precision to place reasonable constraints on the magnification bias $s$.
According to \cite{Lorenz:2017iez}, we need to know $s(z)$ within $5-10\%$ accuracy to avoid systematic parameter biases.

Finally, we computed the combined uncertainty $\sigma(\varepsilon_{\rm P})$ on the  relativistic potential effects. Even the full multi-tracer combination is unable to achieve $\sigma(\varepsilon_{\rm P})<1$. In other words, the potential contribution to the power spectrum is not detectable (signal-to-noise $<1$).

\begin{table}
\caption{Fully marginal uncertainties on $\fnl$ and relativistic effects $\varepsilon_I$. The same as \autoref{tab:err_mar_CP}, except that $\sigma(\fnl)$ and each $\sigma(\varepsilon_I)$ are computed by marginalising over {\em all} other parameters, not just the standard cosmological parameters.}
\label{tab:err_mar_all}
\centerline
{
\begin{tabular}{clccccc}
\\ \hline\hline
Redshift & Survey & {$\sigma(\fnl)$} &  {$\sigma(\varepsilon_{\rm D})$}  & {$\sigma(\varepsilon_{\rm L})$} & $\sigma(\varepsilon_{\rm P})$\\
\hline
$0.1-0.58$ & {BGS} & 45.15  & 8.70  & 0.40  & 60.07\\
& {IM2}  & 44.64  & 26.18  & $-$  & 303.32\\
& {IM2$\otimes$BGS}  & 2.71  & 0.14  & 0.13  & 8.87 \\
\hline
$0.9-1.8$ & {H$\alpha$}  & 30.73  & 26.9  & 0.05  & 15.51 \\ 
$0.35-3.05$ & {IM1} & 6.77  & 9.52  & $-$  & 21.24  \\
$0.6-3.05$ & {IM1$\otimes$H$\alpha$} & 4.37  & 0.37  & 0.03  & 6.78 \\
\hline
$0.1-3.05$ & {(IM2$\otimes$BGS)$\oplus$(IM1$\otimes$H$\alpha$)} & 1.71  & 0.13  & 0.03  & 3.84 \\
& IM2$\otimes$BGS$\otimes$IM1$\otimes$H$\alpha$ & 1.55  & 0.13  & 0.03  & 3.51\\
\hline\hline
\end{tabular}
}
\end{table}

\autoref{tab:err_mar_all}  presents a variation of
\autoref{tab:err_mar_CP} in which we include marginalisation over  $\fnl$ and the relativistic parameters $\varepsilon_I$. As mentioned earlier, this incorporates the  uncertainties on the astrophysical parameters $s$ and $b_{\rm e}$, in addition to the cosmological parameters. 
Comparing  \autoref{tab:err_mar_CP} with \autoref{tab:err_mar_all}, we see that for individual surveys the constraints can be substantially degraded, especially for higher redshift surveys. One exception is lensing which is only slightly affected in single tracer.
When we combine all four surveys, the constraints become insensitive to the additional marginalisation.

\begin{figure}
 \centering
  \includegraphics[width=\textwidth]{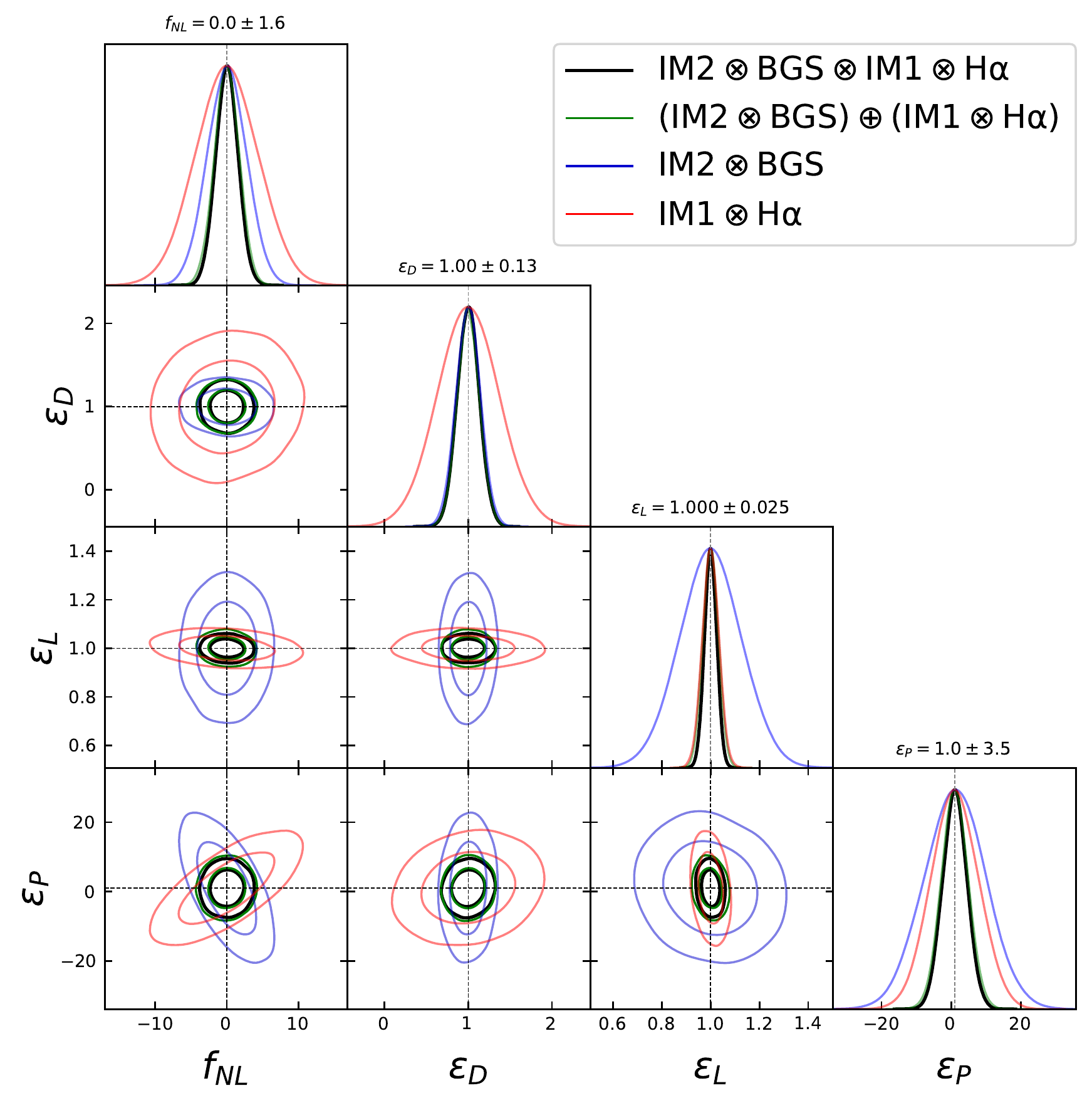}
  \caption{Contour plots of $\fnl$ and relativistic parameters. The correlation of all surveys is shown in \textit{black}, with dotted lines giving the fiducial values. The low-$z$ combination is shown in \textit{blue}, high-$z$  in \textit{red}, and the sum of their Fisher information in \textit{green}. Constraints exclude priors.}
  \label{fig:contour}
\end{figure}

\autoref{fig:contour} displays the  contour plots for the relativistic parameters $\fnl$ and $\varepsilon_I$. The covariances 
include  marginalisation over all parameters in \eqref{pars}, 
as in \autoref{tab:err_mar_all}.  The contours show the total correlated information ({black}), with its low-$z$ ({blue}) and high-$z$ ({red}) components. In {green} we show  constraints from the simple addition of the low-$z$ and high-$z$ Fisher information. 
This is computationally far less expensive, although some information is lost on $\fnl$ {and lensing magnification}. In fact, for relativistic effects, it suffices to consider cross-tracer correlations only in the overlapping footprints. 

We also find that combining low-$z$ and high-$z$ samples breaks degeneracies between $\varepsilon_{\rm D}$ and $\varepsilon_{\rm L}$, and between $\varepsilon_{\rm D}$ and $\varepsilon_{\rm P}$. As an example, the degeneracy between $\varepsilon_{\rm L}$ and $\fnl$ (or $\varepsilon_{\rm D}$) at low-$z$ is orthogonal to the degeneracy at high-$z$. This explains the great improvement seen in  \autoref{tab:err_mar_all} from multi-tracer pairs at low and high redshift, to the full multi-tracer  combination.

In \autoref{tab:err_cond} we give the conditional uncertainties on $\fnl$ and the relativistic parameters to illustrate the robustness of the multi-tracer to  uncertainty in the cosmological model. 
Comparing  \autoref{tab:err_mar_all} and \autoref{tab:err_cond}, it is apparent that the error for single surveys is catastrophically increased by marginalising over the cosmological model, especially at low $z$. The least affected is IM1, due to the high amount of cross-bin correlations included in the analysis. By contrast, the multi-tracer constraints are only slightly improved by not marginalising over cosmological parameters {and clustering bias nuisance parameters}.

\begin{table}
\caption{Similar to  \autoref{tab:err_mar_CP}, except that we display here the {\em conditional} uncertainties on $\fnl$ and relativistic effects $\varepsilon_I$.} 
\label{tab:err_cond}
\centerline
{
\begin{tabular}{clccccc}
\\ \hline\hline
Redshift & Survey & {$\sigma(\fnl)$} &  {$\sigma(\varepsilon_{\rm D})$}  & {$\sigma(\varepsilon_{\rm L})$} & $\sigma(\varepsilon_{\rm P})$\\
\hline
$0.1-0.6$ &{BGS} & 10.56  & 7.41  & 0.39  & 29.79 \\
& {IM2}  & 14.17  & 17.17  & $-$  & 173.93 \\
& {IM2$\otimes$BGS}  & 2.0  & 0.14  & 0.12  & 6.8 \\
\hline
$0.9-1.8$ &{H$\alpha$} & 4.49  & 8.3  & 0.04  & 7.59 \\ 
 $0.35-3.05$ &  {IM1} & 4.42  & 5.52  & $-$  & 9.25 \\
 $0.6-3.05$ & {IM1$\otimes$H$\alpha$} & 2.72  & 0.37  & 0.03  & 4.43 \\
\hline
$0.1-3.05$ &  {(IM2$\otimes$BGS$)\oplus$(IM1$\otimes$H$\alpha$)} & 1.60  & 0.13  & 0.03  & 3.71 \\
& {IM2$\otimes$BGS$\otimes$IM1$\otimes$H$\alpha$} & 1.47  & 0.13  & 0.02  & 3.40\\
\hline\hline
\end{tabular}
}
\end{table}

\subsection*{Systematics}

In reality, we cannot perfectly extract all the observable scales from the data~-- there will be a loss of ultra-large-scale modes due to systematics, e.g., extinction due to Galactic dust or stellar contamination in galaxy surveys, or foreground contamination of intensity mapping. The cosmological signal from  21cm emission is several orders of magnitude lower than the galactic and extra-galactic foreground contamination. In order to extract the cosmological information, it is therefore necessary to first remove or model the systematics in galaxy and intensity mapping surveys. Recent treatments of  ultra-large scale systematics are given in \cite{Rezaie:2021voi} (galaxy survey data) and \cite{Spinelli:2021emp} (21cm intensity mapping simulations). In both cases, information is lost on the largest scales, but the loss is more severe in intensity mapping.

\begin{figure}
\centering
\includegraphics[width=0.49\textwidth]{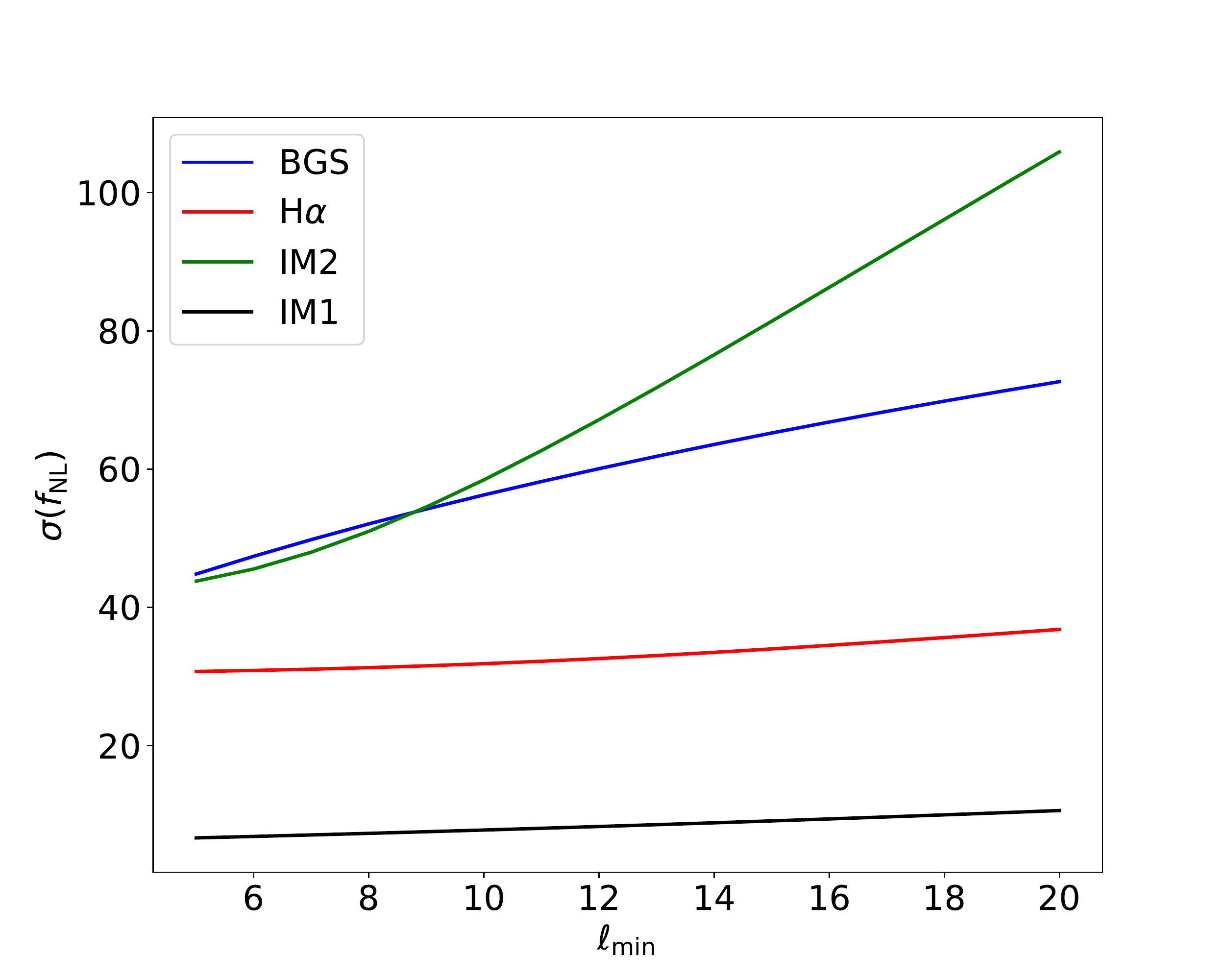}
\includegraphics[width=0.49\textwidth]{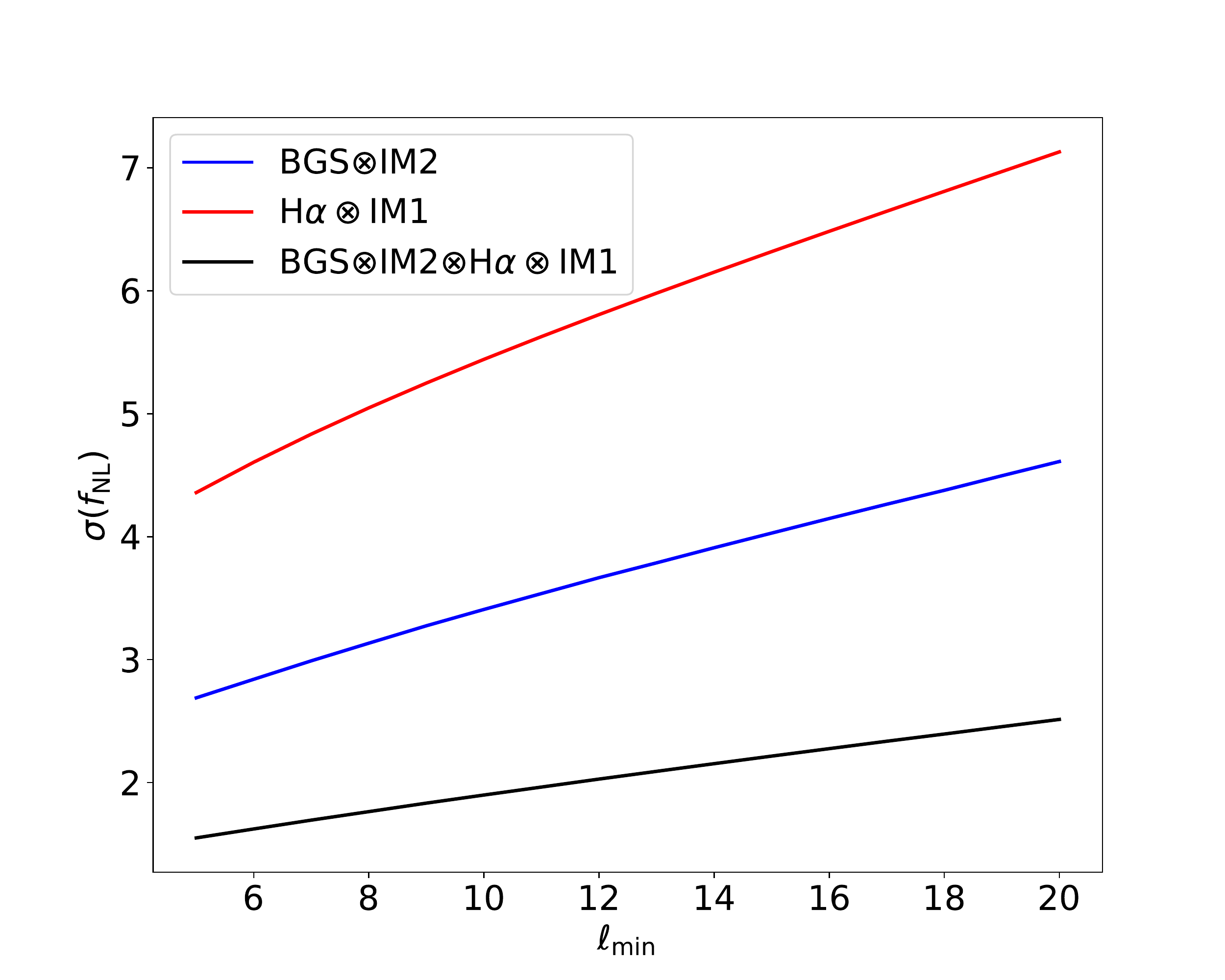}
\includegraphics[width=0.49\textwidth]{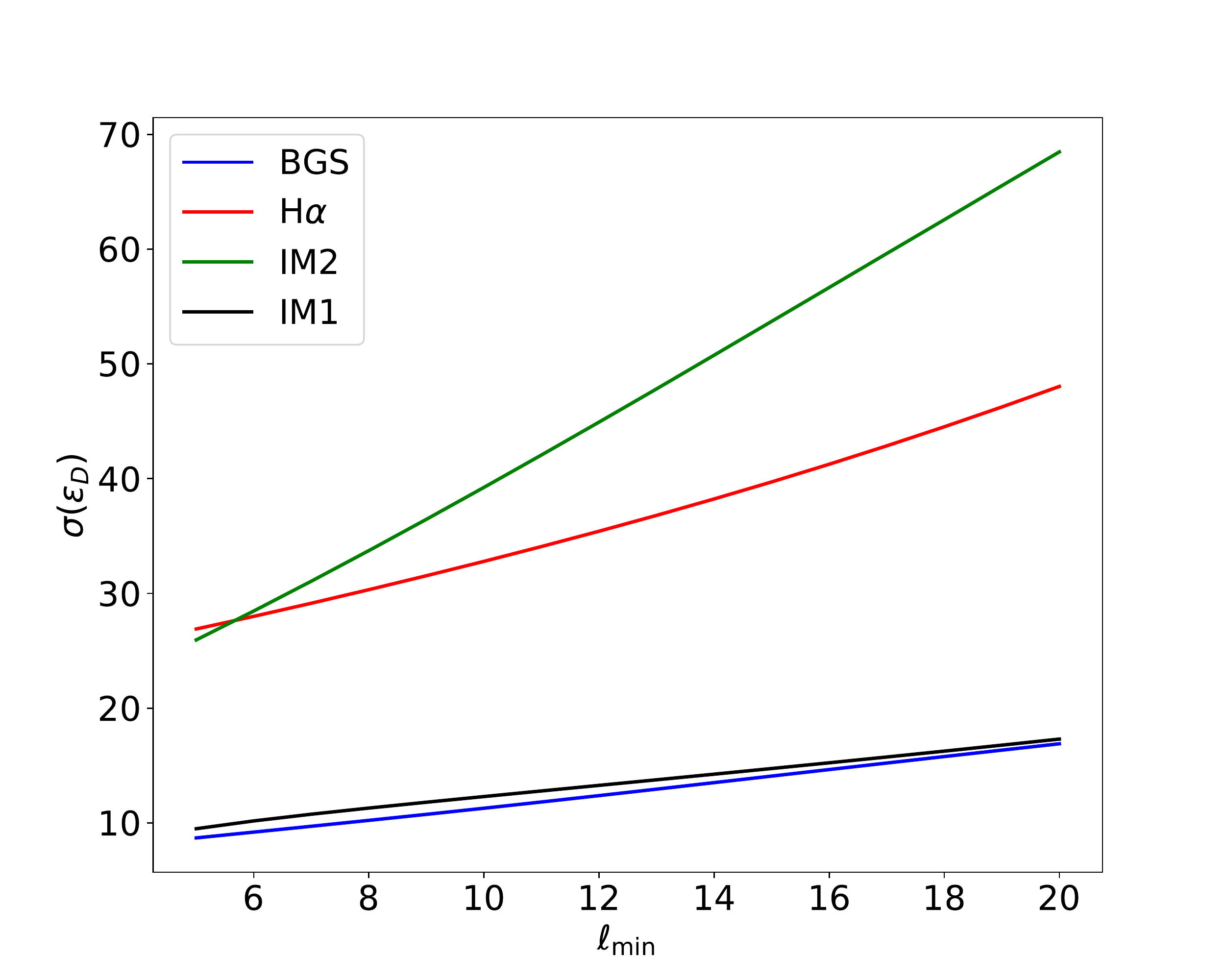}
\includegraphics[width=0.49\textwidth]{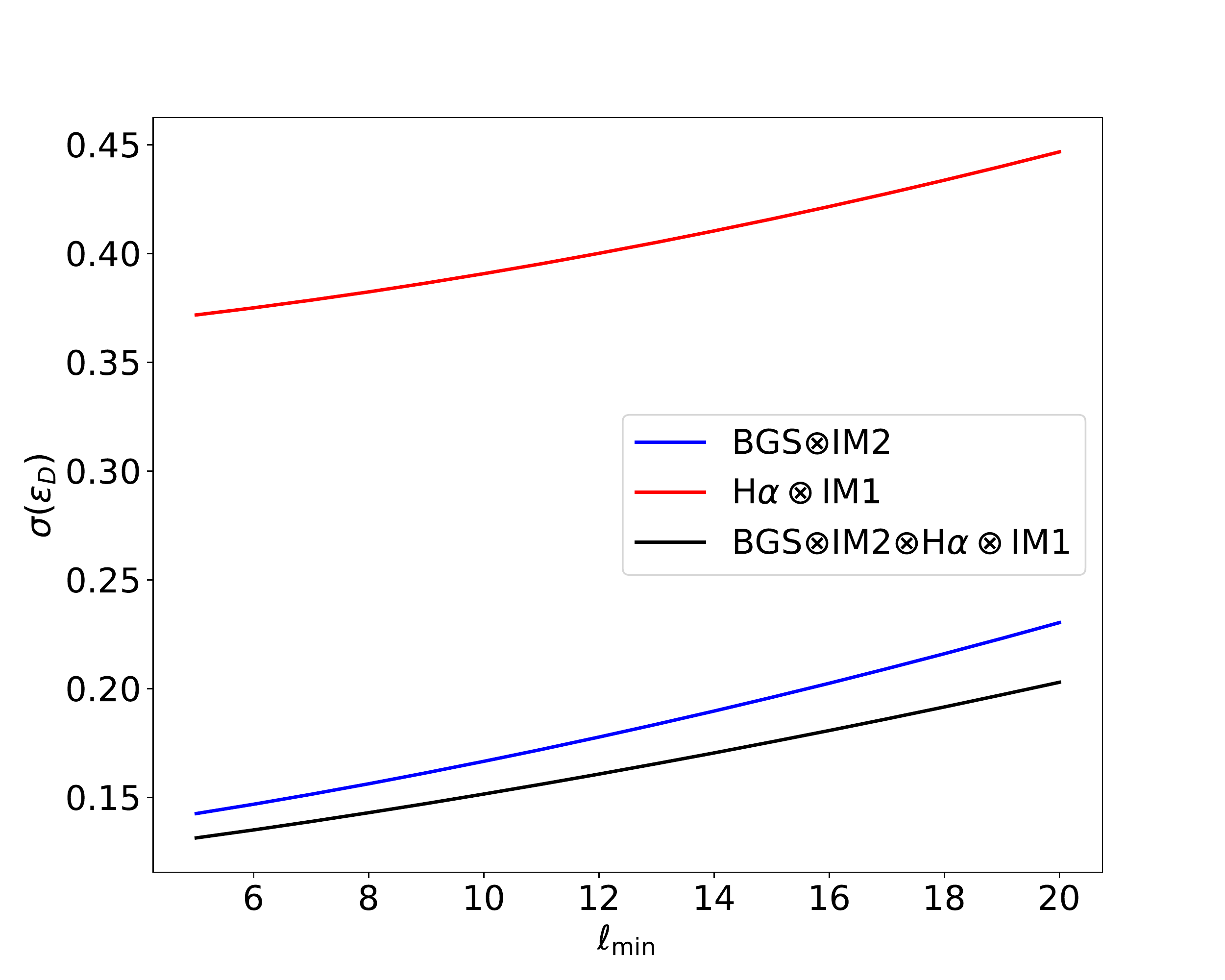}
\includegraphics[width=0.49\textwidth]{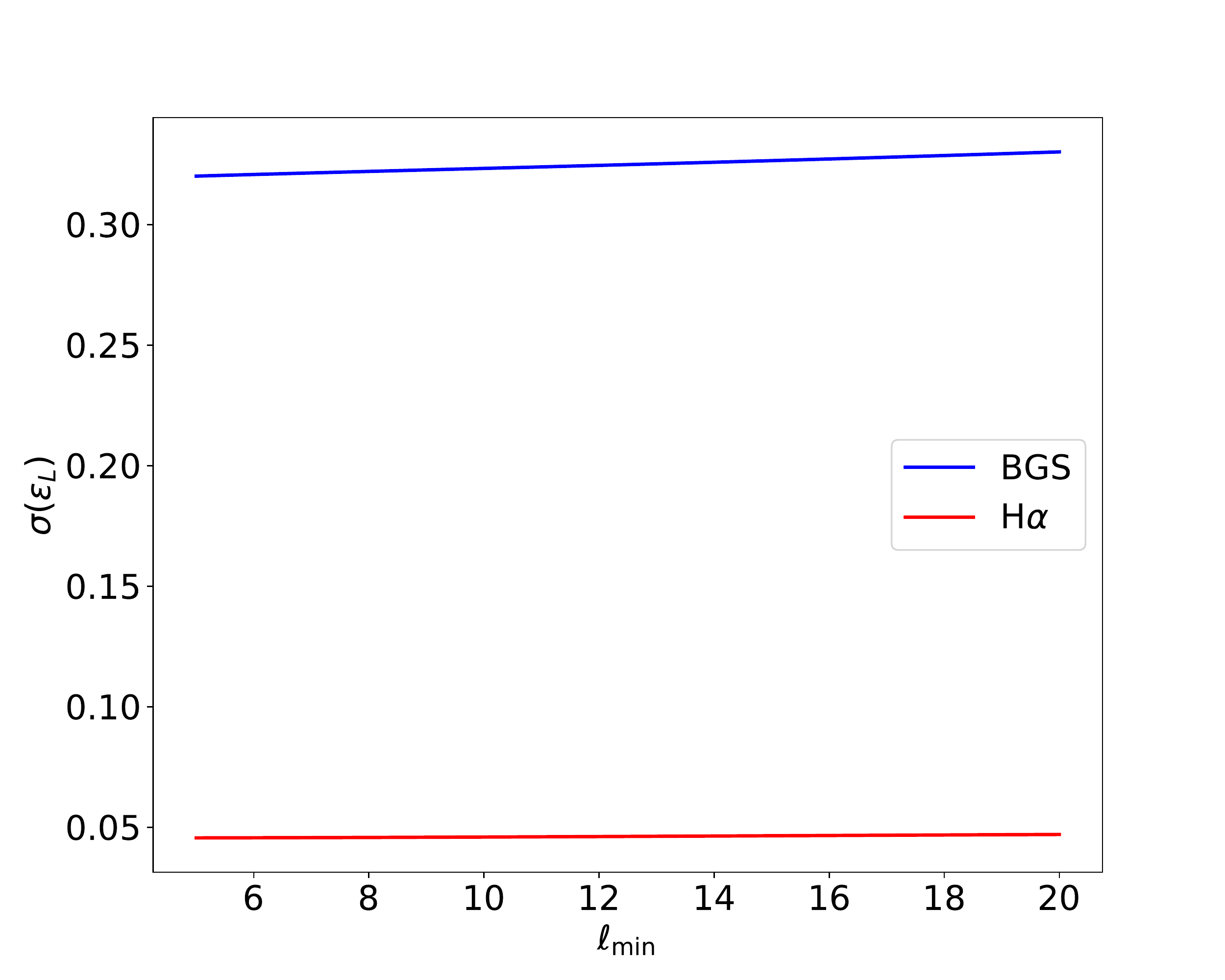}
\includegraphics[width=0.49\textwidth]{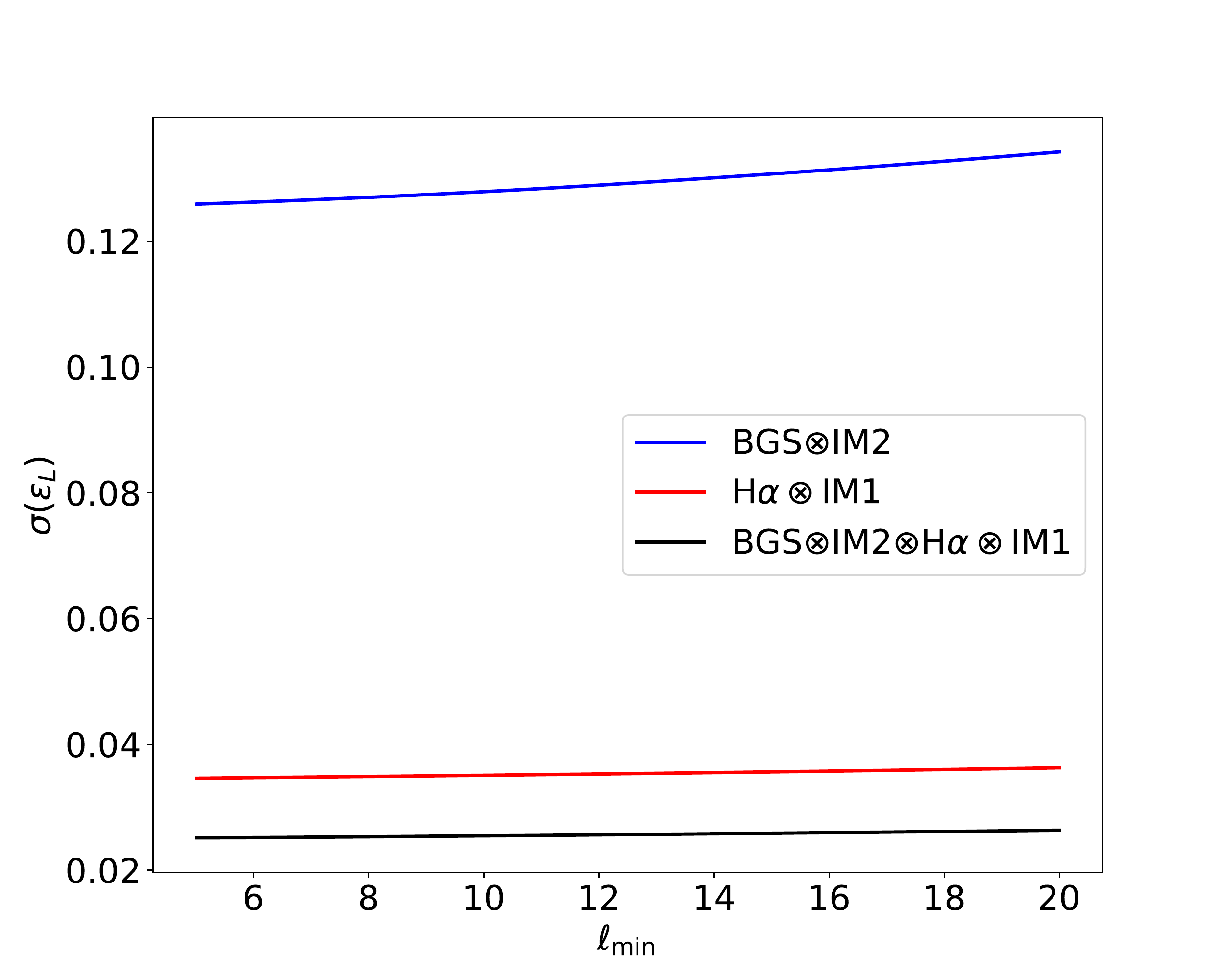}
  \caption{Uncertainty on $\fnl$ (\textit{top}), $\varepsilon_{\rm D}$ (\textit{middle}), and $\varepsilon_{\rm L}$ (\textit{bottom}) as a function of the minimum angular multipole $\ell_{\rm min}$, in the single-tracer (\textit{left panels}) and multi-tracer (\textit{right panels}) cases.  }\label{fig:foreground}
\end{figure}

In order to take account of the loss of some ultra-large scale signal due to systematics, we need to impose a minimum angular multipole $\ell_{\rm min}$. For a multi-tracer analysis, the same $\ell_{\rm min}$ is used for each survey.  In the case of very large sky area, $\ell_{\rm min}\sim 5$ may be feasible for intensity mapping \cite{Witzemann:2018cdx}, and we take this as our `optimistic' minimum.
Then we investigate how the constraints on ultra-large-scale parameters are affected by increasing $\ell_{\rm min}$.

The results for the non-Gaussianity and Doppler parameters are shown in \autoref{fig:foreground}.  It is clear that reducing the maximum scale (lowest $\ell$) does significantly affect the uncertainty in single surveys ({left panels}), for both $\fnl$ ({top}) and $\varepsilon_{\rm D}$ ({middle}), as expected. By contrast, since cosmic variance is cancelled in the multi-tracer ({right panels}), the ultra-large scales have a smaller influence on the constraints, as previously highlighted. The constraints are much more robust to the loss of the largest scales. We conclude that our constraints should not be much affected by ultra-large scale systematics when
using the multi-tracer technique. 

It is worth noting that the constraints from the IM1 survey are less susceptible to increasing $\ell_{\rm min}$, because it includes extremely large-scale correlations along the line of sight, up to $(z_i,z_j)=(0.35,3.05)$. In addition, we find that the uncertainty on $\varepsilon_{\rm D}$ from the BGS survey closely follows IM1, because of the effect mentioned earlier: despite the reduction in observed volume at low $z$, the Doppler amplitude $A_{\rm D}$ is much bigger for BGS, and the magnitude of $\bm n\cdot\bm{v}$ is notably larger. The improved noise properties at low $z$ also contributes to the fact that IM2$\otimes$BGS constrains $\varepsilon_{\rm D}$ much better than IM1$\otimes$H$\alpha$. 

In the case of lensing magnification, the signal does not depend significantly on ultra-large scales and we expect little effect from increasing $\ell_{\rm min}$. This expectation is confirmed by \autoref{fig:foreground} (bottom panels).

\section{Conclusion} \label{sec:conclusion}

In this paper, we investigated how combinations of next-generation large-scale structure  surveys in the optical and radio can constrain primordial non-Gaussianity and relativistic effects. Primordial non-Gaussianity is a powerful probe of the very early Universe, since its signal in the power spectrum is preserved on ultra-large scales. The relativistic effects include the contribution  of lensing magnification to the number density contrast, which provides a probe of gravity and matter that is independent of the probes delivered by weak lensing shear surveys. Of the remaining relativistic effects, the Doppler effect is the most important, and provides another new probe of gravity and matter. 

We reviewed the observable angular power spectra $C^{AB}_\ell(z^A_i,z^B_j)$, which naturally include all the relativistic light-cone effects, in particular the lensing magnification and Doppler effects. We chose a pair of spectroscopic surveys at low redshift and another at high redshift, which have the lowest noise in the respective parts of the electromagnetic spectrum, and which cover large areas of the sky. At low $z$ we chose a DESI-like Bright Galaxy Sample  \cite{2016arXiv161100036D} in the optical and an  HI intensity mapping survey with Band 2 of the SKAO MID telescope \cite{Bacon:2018dui}. The  high redshift pair is a Euclid-like H$\alpha$ spectroscopic survey \cite{Euclid:2019clj} and HI IM with Band 1 of SKAO MID \cite{Bacon:2018dui}.

We identified which pairs of correlations are most sensitive to the lensing and Doppler effects, 
using a signal-to-noise estimator $S^{AB}$ in \eqref{eq:SAB} for angular power spectra (similar to the estimator used in the 2-point correlation function by \cite{Jelic-Cizmek:2020pkh}). The results are
 summarised in \autoref{fig:SN_D_fnl}. This shows clearly that pairs $(z^A_i,z^B_{i+1})$ are optimal for the Doppler effect and are central to the results in \cite{Bonvin:2013ogt,Fonseca:2021gve}. The Doppler effect \eqref{eq:DeltA_D} is also more prominent at low redshifts which is broadly explained by the $1/\chi$ in its amplitude and by the growth of the radial peculiar velocity at low $z$. 
 
 On the other hand, the lensing contribution is stronger at higher redshift and is most relevant in pair correlations $(z^A_1,z^A_{n})$, where $n$ is the total number of bins. Although this is  expected, since lensing is a line-of-sight integrated effect, \autoref{fig:SN_D_fnl} shows it in a visually intuitive manner. 
 
\autoref{fig:SN_D_fnl} also revealed a counter-intuitive feature~-- that the cross-correlation between a galaxy survey and a intensity mapping survey enhances the lensing signal, even though the intensity mapping on its own has no lensing signal. {The point is that background galaxies are lensed by foreground intensity.}

We then computed Fisher forecasts of the constraints on $\fnl$ and the relativistic effects, for all single-tracer cases, for multi-tracer pairs at low and high $z$, and for the full multi-tracer combination of all four surveys. As expected, the full combination provides the most stringent constraints on $\fnl$, improving on the state-of-the-art constraint from Planck  \cite{Akrami:2019izv} by a factor of more than 3. The lensing contribution can be detected at $\sim$2\% accuracy, while the Doppler contribution can be detected with a signal-to-noise of $\sim$8.

We presented our forecasts in two scenarios. First, we assumed that the light-cone effects are perfectly modelled, i.e., the magnification and evolution biases are perfectly known. In the second case, we incorporated model uncertainties in the amplitude of the relativistic contribution to the angular power, by marginalising over these contributions. Predictably, the first scenario provides the best possible constraints. However, the second scenario constraints approach those of the first when we consider multi-tracer pairs~-- and even more when we combine all surveys. The multi-tracer shows one of its strengths: robustness against marginalisation. 

 It is notable that for the lensing contribution, most information comes from the H$\alpha$ survey alone. For the Doppler term, most of the constraints come from the low-redshift multi-tracer. For Doppler and lensing, there is sufficient sensitivity to detect their effects in the $C_\ell$. On the other hand, the relativistic potential terms are too small to be detected, even when combining all surveys. 
 
 In the case of $\fnl$, the multi-tracer results with spectroscopic galaxy surveys delivers $\sigma(\fnl)\sim 1.5$. This is not as good as some forecasts with photometric galaxy surveys (see e.g. \cite{Alonso:2015sfa,Bacon:2018dui,Ballardini:2019wxj}). The spectroscopic multi-tracer is not optimised for $\fnl$, and this was not our focus here. On the other hand, the spectroscopic combination delivers a $\sim$10\% detection of the Doppler effect, which is not possible with photometric  surveys. 

Finally, we stress that it is important to join the low and high redshift surveys to break degeneracies between $\fnl$ and the relativistic effects, as seen in  \autoref{fig:contour}. In addition, the multi-tracer technique provides critical robustness against the loss of  information due to  systematics on ultra-large scales, as shown in \autoref{fig:foreground}. 

There remains an obvious question: {\em if we neglect the relativistic effects, what is the impact on the measured best-fit of $\fnl$ and the other cosmological parameters, in single- and multi-tracer?} This is tackled in a companion paper \cite{Viljoen:2021ocx}. 

\[\]
\acknowledgments

 JV and RM acknowledge support from the South African Radio Astronomy Observatory and the National Research Foundation (Grant No. 75415). RM  also acknowledges support from the UK cience \& Technology Facilities Council (STFC)  (Grant No. ST/N000550/1).
 JF acknowledges support from the UK STFC (Grant No. ST/P000592/1). This work made use of the South African Centre for High-Performance Computing, under the project Cosmology with Radio Telescopes, ASTRO-0945.


\bibliographystyle{JHEP}
\bibliography{Bibliography}

\end{document}